\documentclass[onecolumn,authoryear]{els-mrw} 

\usepackage{amsmath,amssymb,amsfonts,amsthm,makeidx,graphicx}
\usepackage{txfonts}
\usepackage{helvet}
\usepackage[normalem]{ulem}
\usepackage{url}
\usepackage{tocbibind}

\begin{document}

\chapter*{The equation of state for neutron stars}
\label{chap1}

\author{Veronica Dexheimer}%

\address[0]{\orgname{Center for Nuclear Research}, \orgdiv{Department of Physics}, \orgaddress{Kent State University, Kent, OH 44242 USA}}

\articletag{Chapter Article tagline: update of previous edition,, reprint..}

\maketitle

\begin{keywords}
Equation of State, Neutron Star, Dense Matter, Strong Interaction, Exotic Matter, QCD Phase Diagram
\end{keywords}

\begin{glossary}[Nomenclature]
\begin{tabular}{@{}lp{34pc}@{}}
QCD & Quantum Chromodynamics\\
EoS &Equation of State\\
TOV &Tolman–Oppenheimer–Volkoff\\
\end{tabular}
\end{glossary}

\begin{abstract}[Abstract]
This chapter is intended as an introduction to dense matter and the equation of state of neutron stars and their mergers. It begins with a brief description of neutron star interiors, followed by a historical overview of the theoretical frameworks used to describe them, focusing on relativistic formalisms, including different degrees of freedom, models, symmetries, and phases. It also provides an overview of our current understanding of dense matter (including theory, experiments, and observations) and discusses the advances we expect to see in the field over the next decade.
\end{abstract}

\renewcommand*\contentsname{\bf Table of Contents\hspace{19.28cm}}

\begin{minipage}{\textwidth}
\tableofcontents
\end{minipage}

\section{The interior of neutron stars}
\label{interiors}

\subsection{Giving birth to asymmetric matter}
\label{1.1}

Neutron stars are born at the end of the lives of massive stars that run out of fuel for nuclear fusion. Stars that begin their protostellar evolution with masses $M\lesssim8~\rm{M}_{\rm{Sun}}$ become white dwarfs, whereas stars born with masses $M\gtrsim25-30~\rm{M}_{\rm{Sun}}$ become black holes. The upper limit depends on the stellar metallically, i.e., the abundance of elements heavier than hydrogen and helium. Above $\sim8~\rm{M}_{\rm{Sun}}$, stars can fuse elements all the way to iron, after which further fusion reactions no longer release energy. For $M\lesssim25-30~\rm{M}_{\rm{Sun}}$, the gravitational collapse that follows, which after bouncing back the inner region gives rise to a shockwave, lasts for only about a minute, culminating in a supernova explosion that releases a huge amount of energy, enough to temporarily outshine a galaxy~(\cite{BranchWheeler2017}). 

What is left behind becomes a neutron star, the smallest type of star there is. One of the most remarkable characteristics of matter in the interior of neutron stars is its extreme isospin asymmetry. Isospin is a quantum number carried by particles, and isospin symmetry refers to particles interacting in the same way under the strong force. Matter that forms atomic nuclei (other than hydrogen) is approximate isospin symmetric, meaning it has about the same number of protons (Z) and neutrons (N), with heavier nuclei being more asymmetric. While light nuclei (with low number of nucleons, $A=Z+N$) are more symmetric and have a charge fraction $Y_Q\equiv Z/A\sim0.5$ (e.g., for helium $Y_Q=2/4=0.5$), typical heavier nuclei like gold have $Y_Q=79/197\sim0.4$, with the lowest charge fraction for a stable nucleus being an isotope of mercury with $Y_Q=80/204=0.39$.

\begin{figure}[t!]
\centering
\includegraphics[width=.8\textwidth]{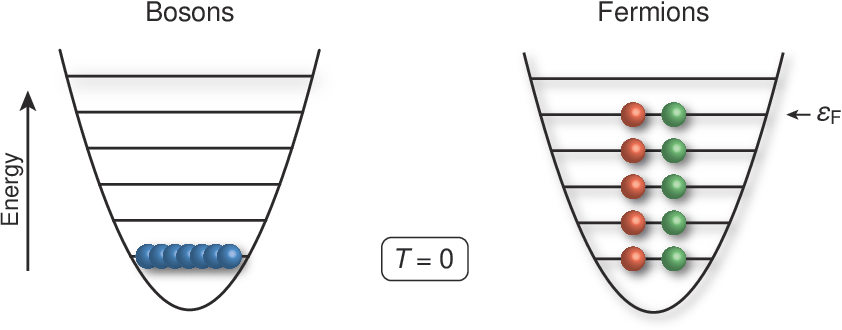}
\caption{Diagram illustrating energy levels for bosons vs. fermions, where in the case of bosons the particles can \emph{condense} into the lowest level. For fermions, the Fermi energy is indicated, the energy level up to which all levels are filled at zero temperature. Figure from~\cite{Will2011InteractingBA}.}
\label{fig0}
\end{figure}

In vacuum, free neutrons have a half-life of about $10$ minutes, $\beta$-decaying into protons, electrons, and anti-neutrinos. Inside a nucleus, however, neutrons are Pauli blocked. This means that in dense environments it is energetically too costly for the neutrons to decay into protons because the Pauli exclusion principle prevents fermions (spin 1/2 particles) from occupying the same energy quantum state, while the low-momentum proton energy states are already filled; see diagram in Fig.~\ref{fig0}. The densities reached during the gravitational collapse of massive stars create a macroscopic environment in which Pauli blocking gives birth to the most isospin-asymmetric matter in the Universe. This happens through a process called \emph{neutronization}, by which a proton captures an electron producing a neutron (that cannot decay back) and a neutrino. As a result, dense matter in neutron stars can reach $Y_Q\sim0.1$, although the exact number is model dependent.

Neutron stars are fully formed approximately one minute after the supernova collapse starts. They typically have masses between $M\sim1$ and $M\sim2~\rm{M}_{\rm{Sun}}$ (with typical values $\sim1.4~\rm{M}_{\rm{Sun}}$) and radii $R\sim12~\rm{km}$, roughly the size of a medium-sized city. The result is an average density (mass over volume) of $\rho=M/V=M/(\frac{4}{3}\pi R^3)\sim4\times10^{14}~\rm{g/cm}^3$, more than the normal nuclear density for a heavy atomic nucleus, $\sim3\times10^{14}~\rm{g/cm}^3$. For comparison, the density of water is $\rho=1~\rm{g/cm}^3$. Equivalently, one can estimate the total number of nucleons in a neutron star as  $A\sim M/m$ (ignoring gravitational binding), with the mass of a nucleon being on average $m=1.67\times10^{-24}~\rm{g}$, giving $A\sim10^{57}$ nucleons, corresponding to a number density $n\equiv A/V\sim0.2~\rm{fm}^{-3}$ (fm is short for femtometer, $10^{-15}~\rm{m}$). For comparison, using the estimated radius of nucleons as $r=0.84~\rm{fm}$, one nucleon would have a number density of $1/(4/3\pi r^3)=0.4~\rm{fm}^{-3}$ (in this case there would be no empty space between nucleons). This is a rough estimate but it means that there is very little space in between the nucleons in a neutron star, with on average only half being empty space,  unlike normal matter, where most space inside atoms is empty.

\begin{figure}[t!]
\centering
\includegraphics[trim={0 1.81cm 0 0},clip,width=.7\textwidth]{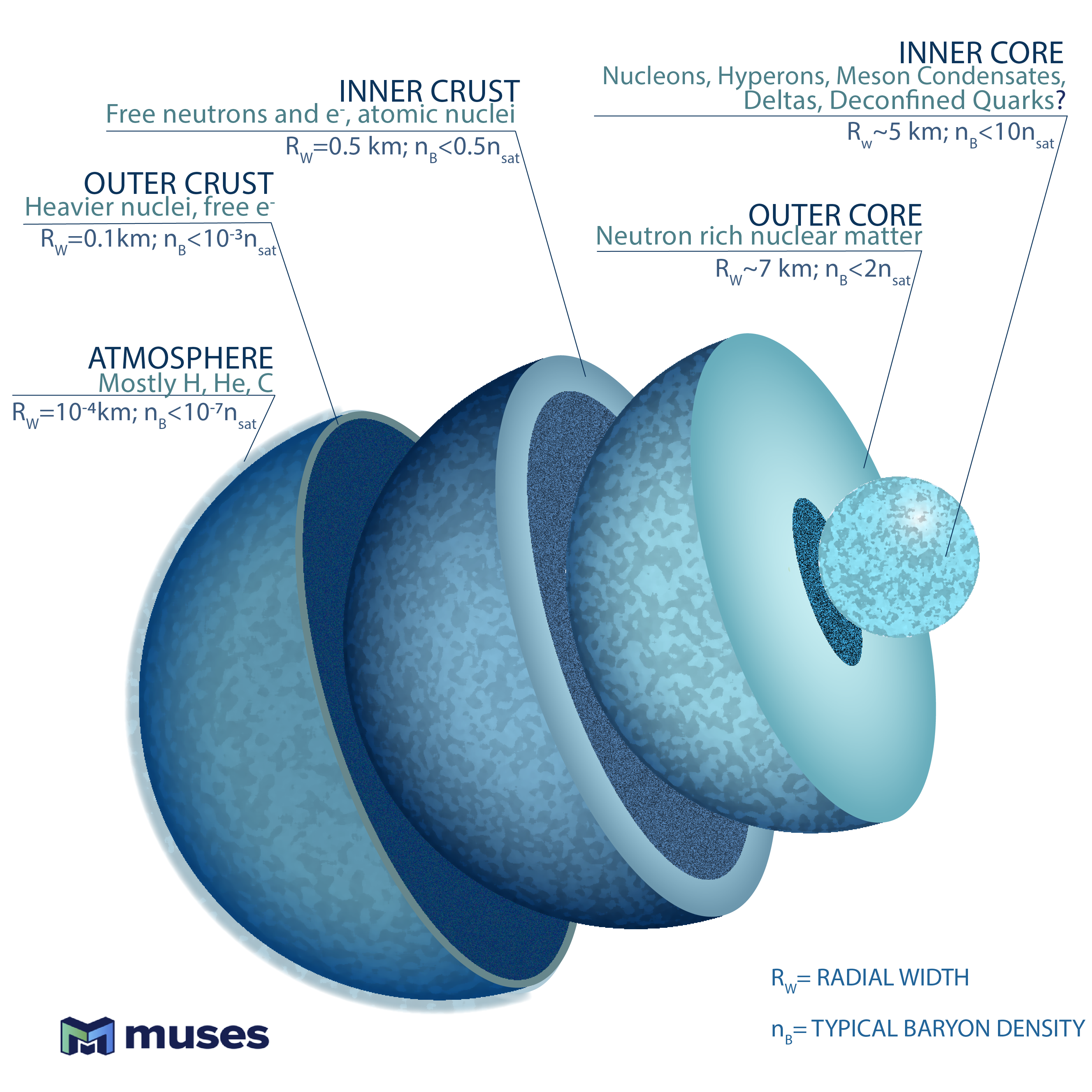}
\caption{Illustration for the interior of a neutron star. We discuss $n_{\rm{sat}}$ in the following, for now, we assume it to be approximately the density of a typical nucleus ($\sim0.12~\rm{fm}^{-3}$). Figure modified from~\cite{MUSES:2023hyz}.}
\label{fig1}
\end{figure}

Although average values are generally informative, gravity, just like on Earth, stratifies matter, increasing the density towards the center and decreasing it towards the surface of approximately spherical objects. Unlike on Earth, gravity is so strong in neutron stars that matter near the surface is organized into a lattice of heavy nuclei (sharing electrons) which we define as \emph{crust}, with neutrons dripping out of the nuclei as the density increases, forming the \emph{inner crust}. As the number density increases further, the nuclei dissolve into uniform matter, referred to as \emph{hadronic matter}, which occupies the region of the neutron star known as the \emph{core}. See Fig.~\ref{fig1} for an illustration of what is expected to be found in the interior of a neutron star. In this figure, both the crust and the core are further divided in two layers and a very thin atmosphere is also depicted. The figure illustrates that the core is much larger than the crust, occupying $\sim90\%$ of the stellar radius and $\sim73\%$ of its volume.

Hadrons are particles that contain quarks. Nucleons (protons and neutrons) are baryons, hadrons made up of 3 valence quarks. Generalizing to different baryons species $i$, we sum over species multiplying the number of particles of each species, $N_i$, by its baryon (quantum) number $B_i$, and redefine the number density of nucleons $n$ to become the baryon (number) density
\begin{align}\label{chap1:eq1}
n_B\equiv \frac{B}{V}= \frac{\sum_{i} B_i N_i}{V} = \sum_{i} B_i n_i\,,
\end{align}
where $n_i=N_i/V$ is the number density for one species and $A$ has been generalized to $B$, the number of baryons. Multiplying the number of particles by $B_i$ allows other particles species to be included in the sum, where $B=1$ for a baryon, $B=-1$ for an antibaryon, $B=1/3$ for a quark, and $B=0$ for other particles. We also generalize the (electric) charge fraction to
\begin{align}\label{chap1:eq2}
Y_Q\equiv \frac{Q}{B}= \frac{\sum_{i} Q_i n_i}{\sum_{i} B_i n_i}= \frac{n_Q}{n_B}\,,
\end{align}
where $Q_i$ is the particle electric charge and $n_Q$ is the charge (number) density. 

However, not all electrons disappear during supernova explosions. Some remain in neutron stars, distributed throughout the space between the nuclei in the crust and the hadronic matter in the core. Electrons are essential for maintaining electric charge neutrality, $n_Q=0$, in the star (remember that  electromagnetism is $\sim10^{36}$ times stronger than gravity). As shown in~\cite{livro:Glendenning}, considering a charged test particle placed on the surface of a charged star with the same sign of electric charge results in a maximum net charge per baryon $Q_{net} \sim 10^{-36}$ (in units of the elementary charge) in the star, beyond which gravity is no longer sufficient to retain it and the test particle is expelled. This simple estimate shows that the net electric charge per baryon must be practically zero in neutron stars.

Shortly after the supernova explosion, neutron stars reach $\beta$-equilibrium, meaning that neutron decay and electron capture balance one another, so that the particle number densities remain constant
\begin{align}\label{chap1:eq3}
n \rightarrow p + e^- + \bar{\nu}_e \quad \text{and} \quad p+ e^- \rightarrow n + \nu_e\,,
\end{align}
with the mean free path of neutrinos being long enough that they can be considered free to leave the star. This can be understood as the neutrino chemical potential (the change in the free energy of the system with respect to the neutrino density) vanishing, $\mu_\nu=0$. The chemical potentials ($\mu$s) of the remaining particles relate through 
\begin{align}\label{chap1:eq4}
\mu_n= \mu_p+ \mu_e^-\,.
\end{align}

This relation can be generalized to other baryons and quarks by expressing their $\mu$s in terms of conserved quantities. In astrophysics, these are usually the baryon number and electric charge
\begin{align}\label{chap1:eq5}
\mu_i= B_i{\mu_B}_i+ Q_i{\mu_Q}_i \,,
\end{align}
where $\mu_B$ is the baryon chemical potential associated with the baryon anti-baryon asymmetry ($\mu_B=0$ corresponds to no asymmetry, as in the matter created in the extremely hot early Universe), while $\mu_Q$ is associated with the isospin asymmetry ($\mu_Q=0$ corresponds to isospin-symmetric matter, as in a typical light nucleus); see~\cite{Aryal:2020ocm} for a detailed discussion of the relationship between electric charge and isospin.
In $\beta$-equilibrium, a similar relation can also be written for the electrons
\begin{align}\label{chap1:eq6}
\mu_e^-= -\mu_Q\,.
\end{align}

Depending on the conserved quantities of the system, Eq.~\eqref{chap1:eq5} can include additional terms, e.g., $S_i{\mu_S}$ whenever strangeness is conserved, as in the case in heavy-ion collisions, which take place on a very short time scale, $t\sim10^{-20}~\rm{s}$, even compared with the timescale of weak interactions, $t\sim10^{-10}~\rm{s}$.

\subsection{Nuclear properties}
\label{1.2}

The unprecedented densities found in the interiors of neutron stars significantly deform spacetime, therefore, any macroscopic description of their properties must rely on general relativity. This includes calculations of stellar masses and radii. On the other hand, microscopic properties of neutron stars, such as the local density, can be calculated without resorting to general relativity because the spacetime can be considered locally flat, with metric variations across fm of only $\sim10^{-19}$~(\cite{livro:Glendenning}).

As a result, neutron stars can be modeled as giant nuclei with $B\sim10^{57}$, with the difference that they are highly isospin asymmetric (e.g., many more neutrons than protons), therefore, the name \emph{neutron stars}. To describe such a giant nucleus, it is reasonable to approximate matter as infinite, thereby neglecting finite-size and surface effects in the core of neutron stars.
Atomic nuclei, on the other hand, are very small. Nucleons on the surface of a nucleus are attracted by fewer neighboring nucleons, and the strong force, being short-ranged, cannot bind nuclei beyond a certain size (208 nucleons in lead, the stable nucleus with most nucleons) because of the Coulomb repulsion between protons.

However, the strong nuclear force is not always attractive. At, very short distances (corresponding to large densities), it becomes repulsive, creating a minimum in the binding energy per nucleon at a density known as the \emph{saturation density}, $n_{\rm{sat}}$ This corresponds to $\rho_B=1.67-1.90\times10^{15}~\rm{g/cm}^3$ or, equivalently, $n_B=0.15-0.17~\rm{fm}^{-3}$~(\cite{1981A&A...102..299H,Gross-Boelting:1998qhi}), a little bit higher than the density of a typical nucleus. At larger densities, atomic nuclei eventually become unstable. In neutron stars, however, this instability corresponds to the dissolution of nuclei into hadronic matter. Around $n_{\rm{sat}}$, there is also strong competition between the nuclear and Coulomb forces, causing nuclei to form larger, non-spherical structures in neutron stars collectively known as \emph{nuclear pasta}~(\cite{Ravenhall:1983uh}). Around $n_{\rm{sat}}$, the properties of nuclei can be measured in laboratory experiments. We discuss the constraints derived from those experiments in Subsection~\ref{3.3}. Beyond $n_{\rm{sat}}$, however, our understanding of matter becomes much more limited, even for the isospin symmetric case. 

\begin{figure}[t!]
\centering
\includegraphics[width=\textwidth]{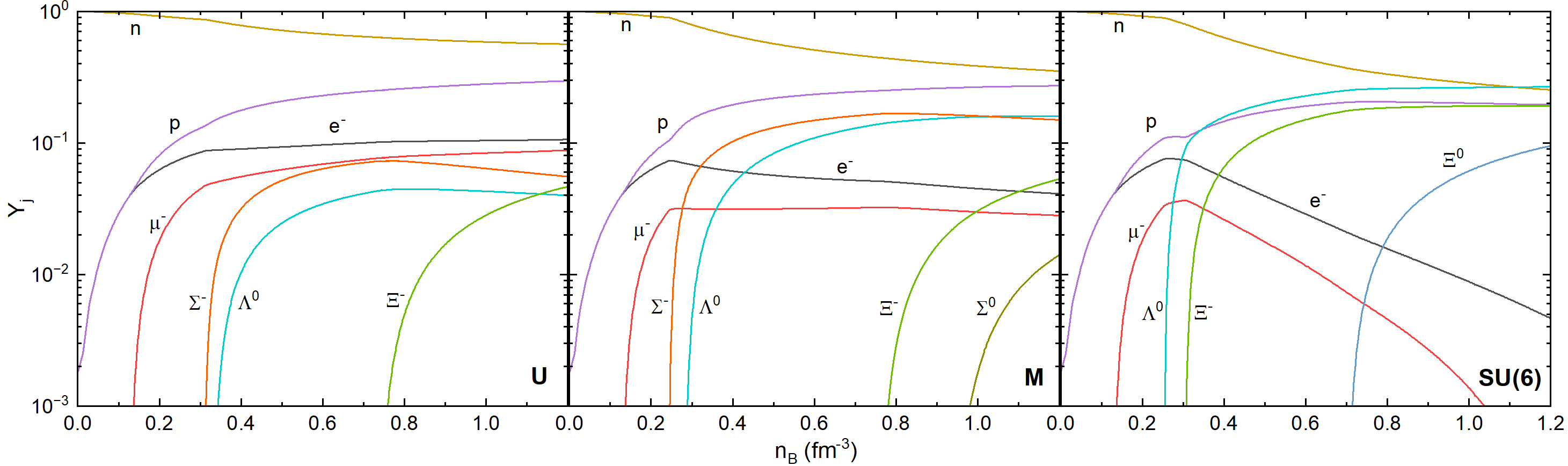}
\caption{MBF model results for particle fractions as a function of density for neutron-star matter. Going right towards increasing baryon density corresponds to going towards the center of a neutron star, although the correspondence is not one-to-one. The different panels correspond to different hyperon coupling schemes. Figure adapted from~\cite{Jacobsen:2026yle}.}
\label{fig2}
\end{figure}

Highly isospin-asymmetric matter is not bound at any density because it is stiffer (i.e., it has a larger pressure). This is a consequence of the Pauli exclusion principle, which forces neutron-rich matter to occupy higher-energy states. This effect becomes particularly relevant for the description of neutron-star cores, as discussed in the following. Heavier hadrons than the nucleons can also appear and remain in $\beta$ equilibrium in neutron stars around $n_B=2\,n_{\rm{sat}}$ and beyond. Hyperons (baryons that contain at least one strange quark) are continuously produced when cosmic rays interact with Earth's atmosphere, but decay rapidly, with lifetimes of $\sim10^{-10}~\rm{s}$ or less. They are also produced in laboratory collisions, where some of their properties can be measured. Nevertheless, isospin asymmetry favors the appearance of hyperons, making neutron stars ideal laboratories for hyperons, which cannot decay back into nucleons because of Pauli blocking. The abundance of hyperons at a given density depends on the their masses, electric charges, and interactions. This can be seen by comparing the different panels of Fig.~\ref{fig2}.

Fig.~\ref{fig2} shows the particle fractions, $Y_i=n_i/n_B$, as a function of baryon density for charge neutral, $\beta$-equilibrated matter as in the core of neutron stars, calculated using the Many-Body Force (MBF) model, which will be briefly discussed in Subsection~\ref{2.2}. The difference between the panels arise exclusively from the different hyperon coupling schemes: in the left panel the hyperons couple in exactly the same way as the nucleons (U - Universal Coupling), in the middle panel the couplings depend on the quark content of nucleons and hyperons (M - Moszkowski Coupling), and in the right panel the couplings account for the strangeness of and isospin proportion between nucleons and hyperons (SU(6) - Spin-Flavor Symmetry Coupling). The hyperons included are the ones from the baryon octet (spin $1/2$) Lambda $\Lambda$, Sigma $\Sigma^+$ $\Sigma^0$ $\Sigma^-$, and Cascade $\Xi^0$ $\Xi^-$. The muon is a heavier lepton, analogous to the electron, and is included for completeness. The figure illustrates the large uncertainties that still exist in our understanding of matter at high densities, although hyperons have been speculated to appear inside neutron stars for decades~(\cite{1962SvA.....5..601A,Glendenning:1982nc}). Hyperons appear in all three panels of Fig.~\ref{fig2} at $n_B\sim2\,n_{\rm{sat}}$, but which hyperons appear, the order at which they appear, their onset densities, and their abundances all depend strongly on the adopted coupling scheme. See e.g., that on the left panel no hyperon reaches $10\%$ of the particle fraction, while on the right panel there are more $\Lambda$ hyperons than neutrons at high densities.

There is also the possibility that non-strange resonances, heavier version of the nucleons, appear in neutron stars. The most relevant resonance for neutron stars is the $\Delta^-$, which is the lightest one of these states and can help fulfill charge neutrality. The appearance of $\Delta$ resonances in neutron stars has been studied for more than two decades~(\cite{Liu:2004se}), and has been proposed as an alternative~(\cite{Schurhoff:2010ph}) to solving the so called \emph{{hyperon puzzle}}. This puzzle arises because the appearance of additional fermionic species tends to soften matter (i.e., reduce its pressure) by redistributing the particles among the available energy states according to the Pauli exclusion principle. As a consequence, relativistic models that include hyperons have greater difficult reproducing neutron-star masses and radii in agreement with observations~(\cite{Bednarek:2011gd}).

Besides baryons, mesons (whose name means ``intermediate'' in Greek because their masses lie between those of electrons and nucleons) can also appear at high densities when describing matter in the interior of neutron stars. In hadronic models, mesons usually provide an effective way to describe the residual strong interaction between the baryons (instead of gluons), most commonly through the scalar meson $\sigma$ and the vector mesons $\omega$ and $\rho$, which are often treated as constant (mean) fields. In addition, mesons can also be treated as explicit degrees of freedom, just like the baryons, the difference being that they don't depend on $\mu_B$, and, being bosons, their thermal population vanishes at zero temperature, unless they condense. Indeed, electrically charged mesons tend to condense in isospin asymmetric matter (as their $\mu$s depend on $\mu_Q$) and at low temperatures. See~\cite{Migdal:1978az} for an early review on the topic of pion condensation in neutron-star matter. In that review, as well as in many subsequent works, the focus has been on how meson condensation decreases the pressure of matter by allowing all mesons to occupy the lowest energy state, thereby potentially destabilizing neutron stars. Note that pions and kaons are normally unstable, with lifetimes $\sim10^{-10}-10^{-8}~\rm{s}$. Nevertheless, under weak equilibrium they are produced as rapidly as they decay. See~\cite{Schaffner-Bielich:2020psc} for a thorough discussion of the appearance of exotic particles in dense matter.

Baryons can exhibit a phenomenon analogous to meson condensation, in which their attractive interaction binds them into Cooper pairs (as described in the BCS theory of~\cite{Bardeen:1957mv}) below a critical temperature that includes typical neutron-star temperatures, $T\lesssim10^{10}~\rm{K}$ or $1~\rm{MeV}$. The binding energy associated with the pair gives rise to a \emph{gap} in the energy spectrum of the fermions relative to the unpaired states. This phenomenon gives rise to superfluidity and, for charged baryons, superconductivity. An analogous phenomenon also occurs in quark matter, where it is known as \emph{color superconductivity}~(\cite{Alford:2007xm}).

\section{Description of dense matter}
\label{thermodynamics}

\subsection{Historical overview}
\label{2.1}

Neutron stars were first hypothesized by Landau in 1931, before the discovery of the neutron, although his work was published only in the following year~(\cite{Landau:1932uwv}). At that time,~\cite{Chandrasekhar:1931ih} had just derived a maximum mass limit for white dwarfs of $M_0\sim0.91~\rm{M_{\rm{Sun}}}$ (later corrected to be $M_0\sim1.44~\rm{M_{\rm{Sun}}}$) based on the pressure of a relativistic degenerate electron gas. This degeneracy pressure arises from the Pauli exclusion principle, which was proposed in~\cite{Pauli:1925nmn} and was first applied to the description of  white dwarfs in~\cite{Fowler:1926zz}. Discussing the possibility of stars with masses $M>M_0$, Landau wrote: ``{\it{We expect that this must occur when the density of matter becomes so great that atomic nuclei come in close contact, forming one gigantic nucleus}}'' but only if ``{\it{a violation of the law of energy, which law, as Bohr has first pointed out, is no longer valid in the relativistic quantum theory, when the laws of ordinary quantum mechanics break down}}'' (see~\cite{Yakovlev:2012rd} for details). The second statement reflects the fact that, without neutrons, neutron star would make no sense. This is because the Coulomb repulsion in a giant nucleus would be so large (obtained by simply scaling up an approximately isospin-symmetric nucleus) that nothing could hold it together, not even gravity. There must also be a source of pressure capable of supporting the star once nuclear fusion is no longer possible, beyond the one provided by the degeneracy pressure of electrons. 

The following year, after Chadwick discovered the neutron~(\cite{Chadwick:1932wcf}), the existence of neutron stars became a realistic possibility. Soon afterward,~\cite{Baade:1934wuu} proposed that massive stars explode as supernovae, leaving behind stars composed of closely packed neutrons. Although neutron stars were not observed until Jocelyn Bell Burnell's discovery of pulsars many years later~(\cite{Hewish:1968bj}), theoretical models had already been developed. \cite{Oppenheimer:1939ne} were the first to model neutron stars as cold, degenerate neutron Fermi gases in 1939. See~\cite{Schmitt:2010pn} for a detailed discussion of a relativistic (meaning including special relativity) free Fermi gas. 

Being almost 2000 times more massive than the electrons, neutrons need to reach a much higher density than the electrons to become degenerate. Degeneracy occurs when $T$ is much lower than the Fermi energy of the fermion, $T<<E_F=\sqrt{k_F^2+M^2}$ (for a free gas), and their momentum $k$ is comparable to their mass $M$, conditions only found naturally inside compact stars. This calculation predicts neutron-star masses not exceeding $M = 0.71~\rm{M}_{\rm{Sun}}$, corresponding to a radius of $R = 9.5~\rm{km}$. See~\cite{Sagert:2005fw} for more details, as well as mass-radius diagrams. Such a low upper limit was immediately recognized as unphysical, since white dwarfs were already known to reach masses up to $M = 1.44~\rm{M}_{\rm{Sun}}$, and neutron stars were expected to be even more massive. However, neutron stars are not composed exclusively of neutrons. $\beta$-equilibrium requires the presence of a finite number of protons together with an equal number of electrons to maintain charge neutrality. The proton (or electron) abundance is usually expressed in terms of fractions $Y_e=Y_Q$, which is typically around $10\%$ in the stellar core, the exact value depending on density and interactions. However, as discussed above, more isospin-symmetric matter is less stiff, further lowering the neutron-star mass limit, in this case not exceeding $M = 0.70~\rm{M}_{\rm{Sun}}$, corresponding to a radius of $R = 9.24~\rm{km}$.

These neutron-star masses and radii were obtained by solving the Tolman–Oppenheimer–Volkoff (TOV) equations~(\cite{Tolman:1939jz,Oppenheimer:1939ne}). The TOV equations are a simplified form of Einstein equation for general relativity describing the hydrostatic equilibrium of stars for the idealized case of spherically symmetric, static, and isotropic stars. They provide an excellent approximation for isolated stars that neither rotate too rapidly nor possess extremely strong magnetic fields. The main part of TOV is a differential equation that determines how the pressure varies with the stellar radius. Its Newtonian counterpart (obtained by neglecting the three general-relativistic terms) simply balances the gravitational force pulling a stellar layer towards the center, against the pressure gradient pushing it outward.  The gravitational and pressure forces exerted by the material outside the shell separately cancel because of spherical symmetry. In both Newtonian and TOV formulations, the hydrostatic equilibrium equation must be integrated throughout the star, starting from a chosen central condition (usually chosen in terms of energy density, $\varepsilon$) out to the stellar surface, where the pressure vanishes. The relation between pressure and energy density (or an equivalent thermodynamic variable) is known as the \emph{equation of state} (EoS).

The EoS varies with baryon density, which increases towards the center of neutron stars; see Fig.~\ref{fig2}. It also depends on the assumed composition, where the inclusion of more exotic particles (anything beyond neutrons, protons, and electrons) further modifies the macroscopic stellar properties. However, the most important ingredient of a neutron-star EoS is the description of the nuclear interactions. The reason why free Fermi gases, \emph{regardless of the composition}, cannot reproduce observed neutron star masses is that the strong interaction provides a dominant contribution to the EoS. This is evident from the fact that observed neutron stars have masses between approximately $1$ and $2~\rm{M}_{\rm{Sun}}$, while the free Fermi gas predicts maximum masses well below this range.

The first attempts to model the strong force relativistically were motivated by the Yukawa potential~(\cite{1955PThPS...1....1Y})
\begin{align}\label{chap2:eq7}
V\propto-\frac{e^{-r}}{r}\,.
\end{align}
This potential is analogous to the Coulomb potential, which decreases linearly with distance, $-1/r$, but the Yukawa potential decreases exponentially at large distances because it arises from the exchange of massive mesons, in analogy with the photon, whose exchange mediates the Coulomb interaction. The two terms required to describe the strong interaction represent attraction and repulsion
\begin{align}\label{chap2:eq8}
V_{\rm{eff}}= -\frac{{g_\sigma,i}^2}{4\pi}\frac{e^{-m_\sigma r}}{r}+\frac{{g_\omega}^2}{4\pi}\frac{e^{-m_\omega r}}{r}\,.
\end{align}
In this picture, the attraction between nucleons is mediated by the $\sigma$ field, which does not correspond to an experimentally observed  meson, but instead represents the residual strong interaction generated by gluons.  The repulsive interaction is mediated by the $\omega$ field, which or may not correspond directly to the physical meson. Here $m_\sigma$ and $m_\omega$ denote the meson masses, which control the interaction range, while $g_{\sigma,i}$ and $g_{\omega,i}$ denote their couplings to a baryon species $i$, which together determine the balance between attraction and repulsion. In natural units, which are used throughout this chapter, the units of mass and distance cancel out.

Building on the earlier non-relativistic work of~\cite{Johnson:1955zz}, both Durr~(\cite{Duerr:1956zz}) and Marx~(\cite{PILKUHN1973460}) developed partially relativistic formalisms to describe nuclear matter inside nuclei, successfully reproducing nuclear saturation with the expected binding energy. Johnson and Teller proposed a linear interaction between baryons and a scalar meson, $g|\psi|^2\phi$, where $\psi$ denotes the nucleon field, $\phi$ the meson field, and $g$ is the coupling strength. Durr was the first to describe  nucleons within this framework using the Dirac equation (thereby naturally including anti-nucleons, the corresponding antiparticles of the nucleons), adopting Johnson and Teller's interaction, while also extending the model to include a repulsive vector meson, $\omega$ (already implemented in a non-relativistic formalism in~\cite{Zeldovich:1961sbr}). 
Further discussion on the role of isospin in nuclear saturation appeared in~\cite{Lee:1974jk} (for a non-relativistic model) and~\cite{KALLMAN1975178} for a relativistic model.

\subsection{Relativistic description}
\label{2.2}

The first fully relativistic model was constructed by~\cite{Walecka:1974qa}, who adopted a covariant formulation, meaning that the theory is independent of the choice of reference frame and, therefore, valid in any coordinate system. He described the nucleons
using a Dirac Lagrangian density, which was later generalized to describe additional baryonic species. The Dirac term contains both kinetic and mass contributions for the baryons. In relativistic quantum field theory, this is the standard description for spin 1/2 fermions (although it is sometimes also used for spin 3/2 fermions). Unlike Schrodinger's theory, in Dirac's theory space and time are treated on an equal footing, while both spin and antiparticles arise naturally within the formalism. The covariant and contravariant derivatives are written as $\partial_\mu=(\partial/\partial t,\nabla)$ and $\partial^\mu=(\partial/\partial t,-\nabla)$, respectively, following the signs of the 
Minkowski's metric
\begin{align}\label{chap2:eq9}
g_{\mu\nu} = \left(\begin{array}{cccc}
1&0&0&0\\0&-1&0&0\\0&0&-1&0\\0&0&0&-1
\end {array}\right)\,.
\end{align}
The $\sigma$ meson is described by the Klein-Gordon Lagrangian density (since it is a scalar meson with zero spin), while the $\omega$ meson is described by the Proca Lagrangian density (since it is a vector meson with nonzero spin, analogous to the photon, but with non-zero mass). The Klein-Gordon and Proca Lagrangians also contain kinetic and mass terms. Finally, interaction terms couple the baryons to both meson fields.

The Walecka Lagrangian density, generalized to include different spin 1/2 baryons~(\cite{Garpman:1979uz}), reads:
\begin{align}\label{chap2:eq10}
\mathcal{L}=\sum_{i=\rm{baryons}}\bar{\psi_i}\left(i\gamma_{\mu}\partial^{\mu}-M_i \right)\psi_i+
\frac{1}{2}\left(\partial_{\mu}\sigma\partial^{\mu}\sigma-m_
{\sigma}^2\sigma^2 \right)+\left(-\frac{1}{4}F_{\mu\nu}F^{\mu\nu}+
\frac{1}{2}m_{\omega}^2\omega_{\mu}\omega^{\mu} \right)+
\sum_{i=\rm{baryons}}g_{\sigma,i}\bar{\psi_i}\psi_i\sigma-\sum_{i=\rm{baryons}}g_{\omega,i}\bar{\psi_i}\gamma_{\mu}\psi_i
\omega^{\mu}\,,
\end{align}
with gamma matrices $\gamma_\mu$ (related to the Pauli matrices) and the field tensor 
$
F^{\mu\nu}=\partial^{\mu}\omega^{\nu}-\partial^{\nu}\omega^{\mu}
$.

It is important to note that Eq.~\eqref{chap2:eq10} does not distinguish between different isospin states (e.g., a proton from a neutron). The couplings $g_{\sigma_i}$ and $g_{\omega_i}$ are the same for both nucleons and are determined by fitting nuclear saturation properties, typically the saturation density and binding energy of isospin-symmetric matter at saturation. Around saturation density, no additional baryon species are expected to be populated. To account for isospin asymmetry, as required for example to describe neutron stars, one would introduce a term like (see Chapter 4 of~\cite{livro:Glendenning} for derivation)
\begin{align}\label{chap2:eq11}
+\left(-\frac{1}{4}\boldsymbol{ \rho_{\mu\nu}\rho^{\mu\nu}}+
\frac{1}{2}m_{\rho}^2\rho_{\mu}\rho^{\mu}\right)-\sum_{i=\rm{baryons}}\frac{1}{2}g_{\rho,i
}\bar{\psi}_i\gamma_{\mu}\psi_i\boldsymbol{\tau\cdot\rho^{\mu}}\,,
\end{align}
with field tensor
$
\boldsymbol{\rho_{\mu\nu}}=\partial_{\mu}\boldsymbol{\rho_{\nu}}-\partial_{\nu}\boldsymbol{\rho_{\mu}}
$.
The bold symbols stand for isovectors  (change sign between members of the isospin multiplet - e.g., proton and neutron) and $\boldsymbol{\tau}$ is made up of the Pauli matrices $\boldsymbol{\tau}=\tau_x \boldsymbol{i}+\tau_y
\boldsymbol{j} +\tau_z \boldsymbol{k}$.

Including the isovector contribution in the Walecka model and rearranging some of the terms gives
\begin{align}\label{chap2:eq12}
\mathcal{L}=\sum_{i=\rm{baryons}}\bar{\psi_i}\left[\gamma_{\mu}(i\partial^{\mu}-g_{\omega,i}\omega^{\mu}
-\frac{1}{2}g_{\rho,i}\boldsymbol{\tau\cdot\rho^{\mu}}
)
-M^*_i \right]\psi_i+
\frac{1}{2}\left(\partial_{\mu}\sigma\partial^{\mu}\sigma-m_
{\sigma}^2\sigma^2 \right)+\left(-\frac{1}{4}F_{\mu\nu}F^{\mu\nu}+
\frac{1}{2}m_{\omega} ^2\omega_{\mu}\omega^{\mu}\right)+\left(-\frac{1}{4}\boldsymbol{ \rho_{\mu\nu}\rho^{\mu\nu}}+
\frac{1}{2}m_{\rho}^2\rho_{\mu}\rho^{\mu}\right)\,,
\end{align}
where the effective mass is defined as
\begin{align}\label{chap2:eq13}
M^*_i=M_i-\sum_{i=\rm{baryons}}g_{\sigma,i}\bar{\psi_i}\psi_i\sigma\,.
\end{align}
In the isospin-symmetric case, the non-relativistic limit for two nucleons reduces to the Yukawa interaction (see~\cite{Walecka:1974qa} for details).

The equations of motion for the baryons and mesons are obtained from the Euler-Lagrange equations, but these equations are highly coupled. A simple way to overcome this difficulty is to apply the mean-field approximation, in which the meson fields are replaced by their mean values and fluctuations around these values are neglected. For the isovector mesons, only the 3rd (neutral) component survives, as it distinguishes the members of the isospin multiplet. Within this approximation, the mesons kinetic terms (that contain derivatives) vanish. The mean-field approximation works remarkably well at high densities and low $T$s, precisely the conditions found in neutron-star interiors. It is worth noting that, within this approximation, the expectation value of the pion field (which has negative intrinsic parity) is zero. Consequently, pions do not appear explicitly in mean-field theories at $T=0$ (unless pion condensation is part of the theory).

\begin{figure}[t!]
\centering
\includegraphics[width=.5\textwidth]{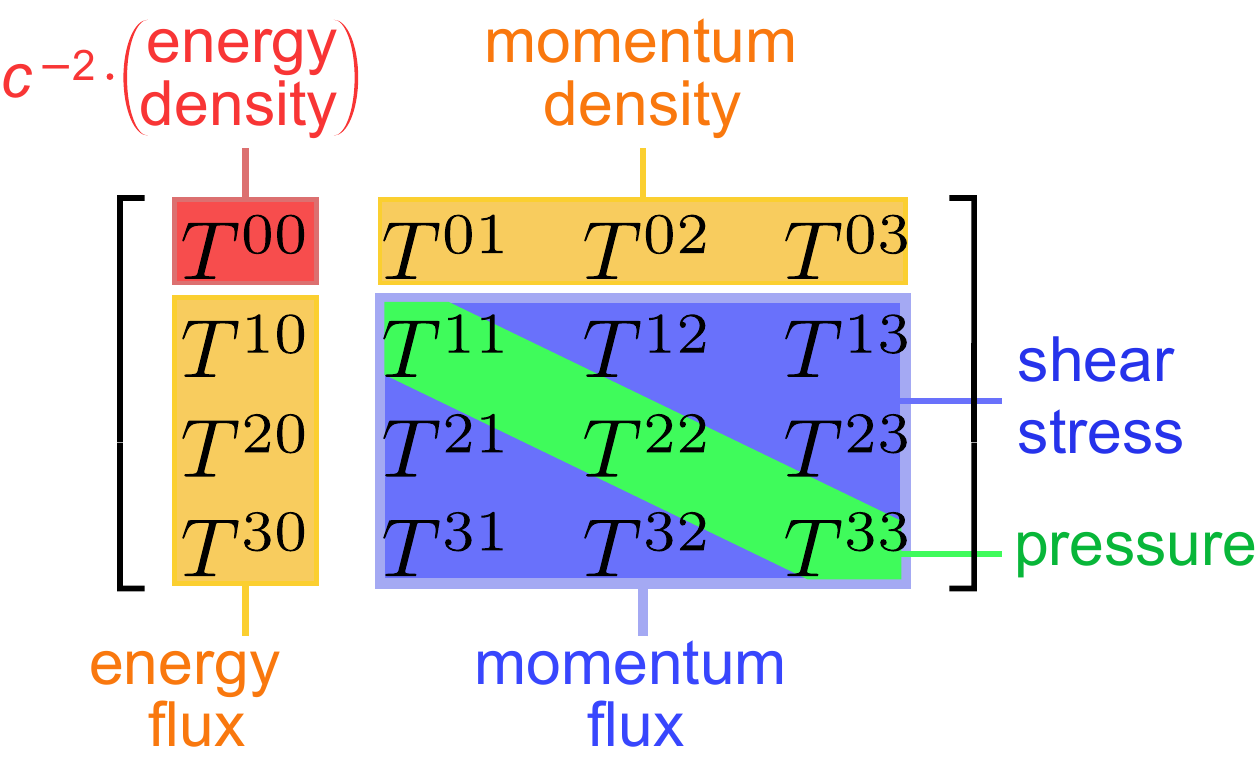}
\caption{Contravariant components of the stress–energy tensor, where the diagonal provides energy density and the different pressure components, while off-diagonal terms provide non-ideal corrections. Figure from Wikipedia.}
\label{fig35}
\end{figure}

Assuming an isotropic ideal fluid (incompressible, irrotational, and with zero viscosity), the energy density and pressure can be obtained from the energy-momentum tensor
\begin{align}\label{chap2:eq14}
T_{\mu\nu}=-\mathcal{L}g_{\mu\nu}+\sum_{\eta}
\frac{\partial\mathcal{L}}{\partial (\partial^{\mu}Q_{\eta})}\ \
\partial_{\nu}Q_{\eta}\,,
\end{align}
where $Q_{\eta}$ denotes either a meson or baryon field,
yielding the energy density, $\varepsilon=T_{00}$, and the isotropic pressure, $P=\frac{1}{3}\delta_{ij}T_{ij}$; see Fig.~\ref{fig35}. There are cases in which the pressure is anisotropic, e.g., in the presence of a strong magnetic field, where the pressure components present different values in different directions~(\cite{Dexheimer:2012mk}), although such cases are beyond the scope of this chapter.

A free gas of leptons (usually electrons but can also include muons) is added by means of a free Dirac Lagrangian to enforce charge neutrality in neutron stars, where ``free'' means non-interacting with respect to the strong force. Alternatively, the pressure can be obtained from the grand potential of the system (and reconstruct the energy density). We discuss this possibility in the Subsection~\ref{2.3}.

Once the pressure and energy density are known, several relevant thermodynamic quantities can be calculates, such as the speed of sound and the compressibility. The speed of sound is a measure of the stiffness of matter and is defined as 
\begin{align}\label{chap2:eq15}
c_s^2=\frac{dP}{d\varepsilon}\,,
\end{align}
usually at a fixed entropy (i.e., under isentropic conditions). In the $T=0$ limit, fixed entropy and fixed $T$ are equivalent. 

In non-relativistic models, $c_s^2$ cannot only become superluminal but, typically, $c_s^2\rightarrow\infty$ as $n_B\rightarrow\infty$.
In the Walecka model, $c_s^2\rightarrow1$ as the density $n_B\rightarrow\infty$, and no superluminal behavior is expected. In principle, even in relativistic models, $c_s^2$ can surpass the speed of light if sufficiently strong repulsive interactions are introduced. Nevertheless, usually, superluminal $c_s^2$ only appears if additional high-order vector-meson self-interactions are included (e.g., $\omega^4$ or $\omega^6$) because fitting saturation properties generally prevents the baryon-vector interaction from becoming too strong.

On the other hand, the Walecka model predicts a compressibility that is too large. The compressibility (originally termed incompressibility because it measures the resistance of matter to compression) is defined as
\begin{align}\label{chap2:eq16}
K=9\frac{dP}{dn_B}\,.
\end{align}
Once more, in the $T=0$ limit it does not matter if the derivative is taken at fixed entropy or fixed $T$. We review nuclear matter constraints in Subsection~\ref{3.3}, but for now it suffices to say that the compressibility predicted by the Walecka model, $K\sim500~\rm{MeV}$, is approximately twice the value expected for isospin symmetric matter at saturation. See~\cite{Serot:1984ey} for a thorough derivation of the model equations.  

To remedy this problem, extensions of the Walecka model introduced non-linear self-interactions terms for the scalar and vector mesons. The first of these, the Nonlinear Boguta-Bodmer model~(\cite{Boguta:1977xi}), introduces scalar-meson self-interaction terms through the potential
\begin{align}\label{chap1:eq17}
U(\sigma)=\frac 1 3 b\sigma^3+\frac 1 4 c\sigma^4\,,
\end{align}
which significantly reduces the compressibility. \cite{Bodmer:1991hz} later introduced self-interactions for the vector mesons, including a dependence on $Z$ in the coupling. Modern versions of the Walecka model also include explicit vector-isovector terms~(\cite{Mueller:1996pm}), which are important to describe asymmetric nuclear properties~(\cite{Horowitz:2000xj}) and neutron star observations~(\cite{Horowitz:2002mb,Dexheimer:2018dhb}).

A completely different approach to improve the Walecka model consists of replacing the (minimal) Yukawa inspired linear coupling between the scalar meson and the baryons with more general non-linear couplings. The first model of this type is the Zimanyi-Moszkowski model~(\cite{Zimanyi:1990np}), in which a dependence on the $\sigma$ field is introduced into the nucleon kinetic term
\begin{align}\label{chap2:eq18}
\mathcal{L}=\left(\frac{g_{\sigma,i}\sigma}{M_i}\right)\bar{\psi_i}i\gamma_{\mu}\partial^{\mu}\psi_i\,,
\end{align}
and, in some formulations, also into the vector coupling and the effective nucleon mass. A subsequent generalization that introduced arbitrary powers of the term $\left(1+\frac{g_{\sigma,i}\sigma}{M_i}\right)^{-1}$ in couplings was first suggested in~\cite{Delfino:1995ea} and then pursued in~\cite{Taurines:2000xz,Taurines:2000zb}, leading to the generalized couplings
\begin{align}\label{chap2:eq19}
g_{\sigma,i}\sigma\bar{\psi_i}\psi_i\rightarrow g^*_{\sigma,i}\sigma\bar{\psi_i}\psi_i =\frac{g_{\sigma,i}\sigma}{\left(1+\frac{g^*_{\sigma,i}\sigma}{\lambda M_i} \right)^\lambda}\bar{\psi_i}\psi_i\,.
\end{align}
This framework is now known as the Many-Body Forces (MBF) model. The name steams from the fact that these non-linear interactions effectively mimic many-body correlations, without introducing them explicitly. This can be understood by expanding the non-linear coupling $g^*$ as a series of meson self-interaction terms, where each term represents a different medium contribution, while the overall strength of the expansion is governed by the parameter $\lambda$. The effective coupling $g^*$ was later generalized to modify the vector-mesons couplings as well~(\cite{Dexheimer:2007mt}) and to depend on additional scalar mesons~(\cite{Gomes:2014aka}).
 
A similar extension of the Walecka model consists of making the couplings \emph{explicitly} dependent on the baryon density. Although in the MBF model, the couplings $g^*$ depend on scalar mesons, which themselves depend on density, this dependence in indirect. Several attempts were made to introduce explicit density dependence into relativistic models, e.g.,~\cite{Haddad:1993zz}. Nevertheless, the Density-Dependent (DD) model~(\cite{Typel:1999yq}) was the first one to include explicit density dependence in a covariant way, while also including the rearrangement terms required to ensure thermodynamical consistency (these terms are not required when the couplings depend on the meson fields instead of density). The relevant terms from the DD Lagrangian density are:
\begin{align}\label{chap2:eq20}
\bar{\psi_i}\Gamma_{\sigma,i}\sigma\psi_i-\bar{\psi_i}\gamma_\mu \Gamma_{\omega,i}\omega^\mu\psi_i-\bar{\psi_i}\gamma_\mu \Gamma_{\rho,i}\frac{\boldsymbol{\tau}}{2}\cdot\boldsymbol{\rho^\mu}\psi_i\,,
\end{align}
with the $\sigma$ and $\omega$ meson couplings chosen to approximate ab-initio calculations of isospin symmetric nuclear matter
\begin{align}\label{chap2:eq21}
\Gamma_{\sigma,\omega,i}=\Gamma_{\sigma,\omega,i}{\big\rvert}_ {n_{\rm{sat}}} a_{\sigma,\omega,i}\frac{1+b_{\sigma,\omega,i}(n_B/n_{\rm{sat}}+d_{\sigma,\omega,i})^2}{1+c_{\sigma,\omega,i}(n_B/n_{\rm{sat}}+d_{\sigma,\omega,i})^2}\,,
\end{align}
with eight (real, positive, not independent) parameters $a_i$, $b_i$, $c_i$, and $d_i$, and the $\rho$ meson coupling being
\begin{align}\label{chap2:eq22}
\Gamma_{\rho,i}=\Gamma_{\rho,i}{\big\rvert}_{n_{\rm{sat}}} e^{-a_{\rho,i}(n_B/n_{\rm{sat}}-1)}\,,
\end{align}
with one parameter $a_i$. 

All the modifications described above improve the agreement between the corresponding relativistic models and experimental and observational constraints. The additional parameters provide greater flexibility when fitting parameters, particularly those related to isospin-asymmetric matter. This is usually performed as a second step, after fitting the isospin-symmetric nuclear properties, such as $n_{\rm{sat}}$, binding energy, compressibility, nucleon effective mass, etc. The second step includes symmetry energy at $n_{\rm{sat}}$ (and possibly also its derivatives as a function of density), as well as neutron-star masses, radii, and tidal deformability. Models with density-dependent couplings, whether implicit or explicit, can additionally reproduce observational constraints with greater independence from the symmetry energy at saturation, although not independently from its density derivatives.

\subsection{Equation of state}
\label{2.3}

Without using the ideal fluid assumption, the EoS can be derived using thermodynamics. In this approach, the derivation depends on the choice of statistical ensemble, which consists of a very large set of ``copies'' of the system corresponding to different microstates, while reproducing the same macroscopic attributes. While in the microcanonical ensemble all members of have the same energy and number of particles, in the canonical ensemble members have only the same number of particles, and in the grand canonical ensemble members have neither constraint. In principle, all statistical ensembles become equivalent in the thermodynamic limit, provided fluctuations can be neglected~(\cite{Satarov:2020loq}). 

Whereas in low-energy nuclear physics the canonical ensemble is usually adopted, in high-energy nuclear physics the grand canonical ensemble is adopted. The grand potential, which is the free energy of the grand canonical ensemble, is the natural quantity for describing matter at finite $T$, when thermal energies become comparable to the particle masses (in practice, when $T$ approaches a few percent of the particle mass). This choice is particularly appropriate because, at finite $T$, the number of particles is no longer conserved, owing to the continuous creation and annihilation of particle-antiparticle pairs. For electrons, this correspond to $\sim10^8~\rm{K}\sim10^{-2}~\rm{MeV}$ (compared to their mass $\sim0.5~\rm{MeV}$), whereas, for nucleons, this corresponds to $\sim10^{11}~\rm{K}\sim20~\rm{MeV}$ (compared to their mass $\sim939~\rm{MeV}$).

Starting from the grand potential density
\begin{align}\label{chap2:eq23}
\Omega/V=-\mathcal{L}- T \sum_i \frac{\gamma_i}{(2\pi)^3}\int_0^\infty d^3k \ln{\left(1+ e^{-\frac{1}{T}(E^*_i\mp\mu^*_i)}\right)}\,,
\end{align}
where the upper sign refers to particles and the lower sign refers to antiparticles, the index $i$ runs over all fermionic species (including baryons, quarks, and leptons), $\gamma_i$ denotes the degeneracy factor (e.g., $\gamma_i=2$ for spin-$1/2$ fermions if the different projections are treated equally and an additional factor of $3$ if the quark colors are treated equally), $k$ is the particle momentum, $E^*$ is the effective energy level $E_i^*=\sqrt{{k_i}^2+{M_i^*}^2}$, $M_i^*$ the effective mass, and in the mean-field approximation relativistic models yield $\mu_i^*=\mu_i-g_{\omega,i}\omega
-\frac{1}{2}g_{\rho,i}\rho$ (which may also include contributions from additional vector meson fields), where $\mu_i$ is defined in Eq.~\eqref{chap1:eq5}. For leptons,  $M_i^*=M_i$, $E_i^*=E_i$, and $\mu_i^*=\mu_i$.

Rewriting the momentum integral in spherical coordinates and assuming spherical symmetry, the angular integration can be carried analytically, reducing the momentum integral to $\int d^3k$ integral as $4\pi \int k^2 dk$. Then, integrating by parts yields
\begin{align}\label{chap2:eq24}
\Omega/V=- \sum_i \frac 1 3 \frac{\gamma_i}{2\pi^2}\int_0^\infty dk \frac{k_i^4}{E_i^*}\left(1+ e^{\frac{1}{T}(E^*_i\mp\mu^*_i)} \right)^{-1}\ - \ {\mathcal{L
}}_{\rm{self~int.}}\ \,,
\end{align}
where the Dirac term in the Lagrangian density vanishes when applying the equation of motion for the fermions. Within the mean-field approximation, the meson kinetic terms vanish, whereas the meson self-interactions, including mass terms, survive. At $T=0$,  the upper integral limit becomes the Fermi momentum, $k_{Fi}$, and $E_{Fi}^*=\mu_i^*$.

The Gibbs-Duhen relation written in terms of a generic conserved charge ``$x$'' is
\begin{align}\label{chap2:eq25}
E=-PV+TS+\mu_xN\,.
\end{align}
In the grand canonical ensemble, this becomes
\begin{align}\label{chap2:eq26}
\Omega=-PV=E-TS-\mu_xN_x\,,
\end{align}
with differential
\begin{align}\label{chap2:eq27}
d\Omega=-PdV-SdT-N_xd\mu_x=0\,,
\end{align}
which is minimized in thermodynamical equilibrium. In the infinite-volume limit, Eq.~\eqref{chap2:eq26} is usually written as
\begin{align}\label{chap2:eq28}
P=-\varepsilon+Ts+\mu_xn_x\,,
\end{align}
where we have defined the following densities $\varepsilon=E/V$, $s=S/V$, and $n_x=N_x/V$. The differential, Eq.~\eqref{chap2:eq27}, then becomes
\begin{align}\label{chap2:eq29}
dP=sdT+n_xd\mu_x=0\,.
\end{align}
From Eq.~\eqref{chap2:eq29} we can write $s=\partial P/\partial T\Big|_{\mu_x}$ and different densities $n_B=\partial P/\partial \mu_B\Big|_{T,\,\mu_{x\neq B}}$, $n_Q=\partial P/\partial \mu_Q\Big|_{T,\,\mu_{x\neq Q}}$, etc. Equivalent expressions for the other commonly used ensembles can be found in Appendix E of~\cite{Cruz-Camacho:2024odu}.

\subsection{Phase stability}
\label{2.4}

Whenever multiple phases are present, or even when only one phase exists, stability must be verified. For example, a model may reproduce only one phase, yet that phase may not be more stable than vacuum, meaning that no real particles (as opposed to virtual particles that also populate vacuum) are produced. To verify stability, we begin with the appropriate free energy (for the chosen ensemble) and verify that it is minimized by requiring its first differential to vanish and its second variation to be positive. In the grand canonical ensemble, in the infinite volume limit, the first condition is already satisfied by Eq.~\eqref{chap2:eq28}, while the second condition can be verified by requiring that the determinant of the Hessian matrix (including permutations)
\begin{align}\label{chap2:eq30}
    M&=\begin{bmatrix}
    \frac{\partial^2 P}{\partial T^2}\Big|_{\mu_x}&
    \frac{\partial^2 P}{\partial T\partial \mu_x}\Big|_{T} \\
    \frac{\partial^2 P}{\partial \mu_x\partial T}\Big|_{\mu_x}&
    \frac{\partial^2 P}{\partial \mu_X^2}\Big|_{T}  \\
    \end{bmatrix}\,,
\\
    M&=\begin{bmatrix} \label{chap2:eq31}
    \frac{\partial s}{\partial T}\Big|_{\mu_x} & 
    \frac{\partial s}{\partial \mu_x}\Big|_{T}  \\
   \frac{\partial n_x}{\partial T}\Big|_{\mu_x} & 
   \frac{\partial n_x}{\partial \mu_x}\Big|_{T} 
    \end{bmatrix}\,,
\end{align}
and its submatrices $\ge0$. At $T=0$, this condition reduces to
\begin{align} \label{chap2:eq32}
    \frac{\partial n_x}{\partial \mu_x} &\ge 0\,,
\end{align}
where it is implied that for, e.g., $x=B$, the other variables (e.g., $x=Q$) are kept constant. This condition is commonly referred to as chemical hardness criterion and its positivity is a requirement by  the Le Chatelier principle for thermodynamic stability~(\cite{chandler1987introduction}).

We can also rewrite Eq.~\eqref{chap2:eq32} as
\begin{align}\label{chap2:eq33}
\frac{\partial n_x}{\partial P}\frac{\partial P}{\partial \mu_x} &\ge 0\,,\\
\frac{\partial n_x}{\partial P}n_x &\ge 0\,,\label{chap2:eq34}
\end{align}
which implies
\begin{align}\label{chap2:eq35}
\frac{\partial P}{\partial n_x}\ge 0\quad \rm{for}\quad n_x\ge0\,,
\end{align}
which can also be obtained within the microcanonical or canonical ensembles (see Appendix E of~\cite{Cruz-Camacho:2024odu} for details). For $x=B$, $\frac{\partial P}{\partial n_B}\ge 0$ is equivalent to requiring a positive compressibility, Eq.~\eqref{chap2:eq15}, also a requirement for stability as part of the Le Chatelier principle. 

For charge neutral, $\beta$-equilibrated matter, Eq.~\eqref{chap2:eq28}, together with Eq.~\eqref{chap1:eq1}, Eq.~\eqref{chap1:eq5}, Eq.~\eqref{chap1:eq6}, and charge neutrality condition, $n_Q=0$, reduces to 
\begin{align}\label{chap2:eq36}
P=-\varepsilon+Ts+\mu_Bn_B\,,
\end{align}
meaning that checking stability with respect to $n_B$ alone is sufficient. Otherwise, one would have to check stability with respect to every conserved charge.

Another necessary condition for stability is that $c_s^2$, Eq.~\eqref{chap2:eq15}, is positive. If any of these conditions is violated, the system will settle into another stable phase (or vacuum if no stable phase exists). If several stable phases coexist, the system will settle in the most stable phase (corresponding to the largest pressure in the infinite-matter limit of the grand canonical potential), while the remaining (locally) stable phases are denoted metastable.

As the density increases and new particle species (e.g., hyperons) appear in the system, new phases of matter may emerge. In most cases, the transitions between these phases are smooth, and, consequently, they are not usually regarded as phase transitions. Formally, the order of a phase transitions is defined based on which order of derivative of the free energy becomes discontinuous. The associated order parameter (a variable that characterizes the different phases) is also discontinuous. For example, a discontinuity in the baryon density (the first derivative of the grand potential or pressure with respect to $\mu$) characterizes a first order phase transition. Phase transitions of order higher than second  are sometimes thought of as crossovers (although, formally, a crossover would not present discontinuities in any order of derivative).

The order of the phase transition associated with the appearance of hyperons is model dependent, and also dependent on, e.g., isospin-asymmetry. For example, it was shown in~\cite{Cruz-Camacho:2024odu} that, within the Chiral Mean Field (CMF) model, a phase with hyperons appears typically as third order, although in a few cases it can instead become a first-order phase transition leading to a highly strange (yet still hadronic) phase, depending on the vector-meson self-interactions terms, as well as $\mu_Q$ and $\mu_S$ values.

Phases with different underlying symmetries are expected to be connected through first-order phase transitions. Two phases transitions that are expected to emerge at high densities, possibly in the inner cores of massive neutron stars, are those associated with chiral symmetry restoration and quark deconfinement.

\subsection{Chiral symmetry}
\label{2.5}

\begin{figure}[t!]
\centering
\includegraphics[width=.4\textwidth]{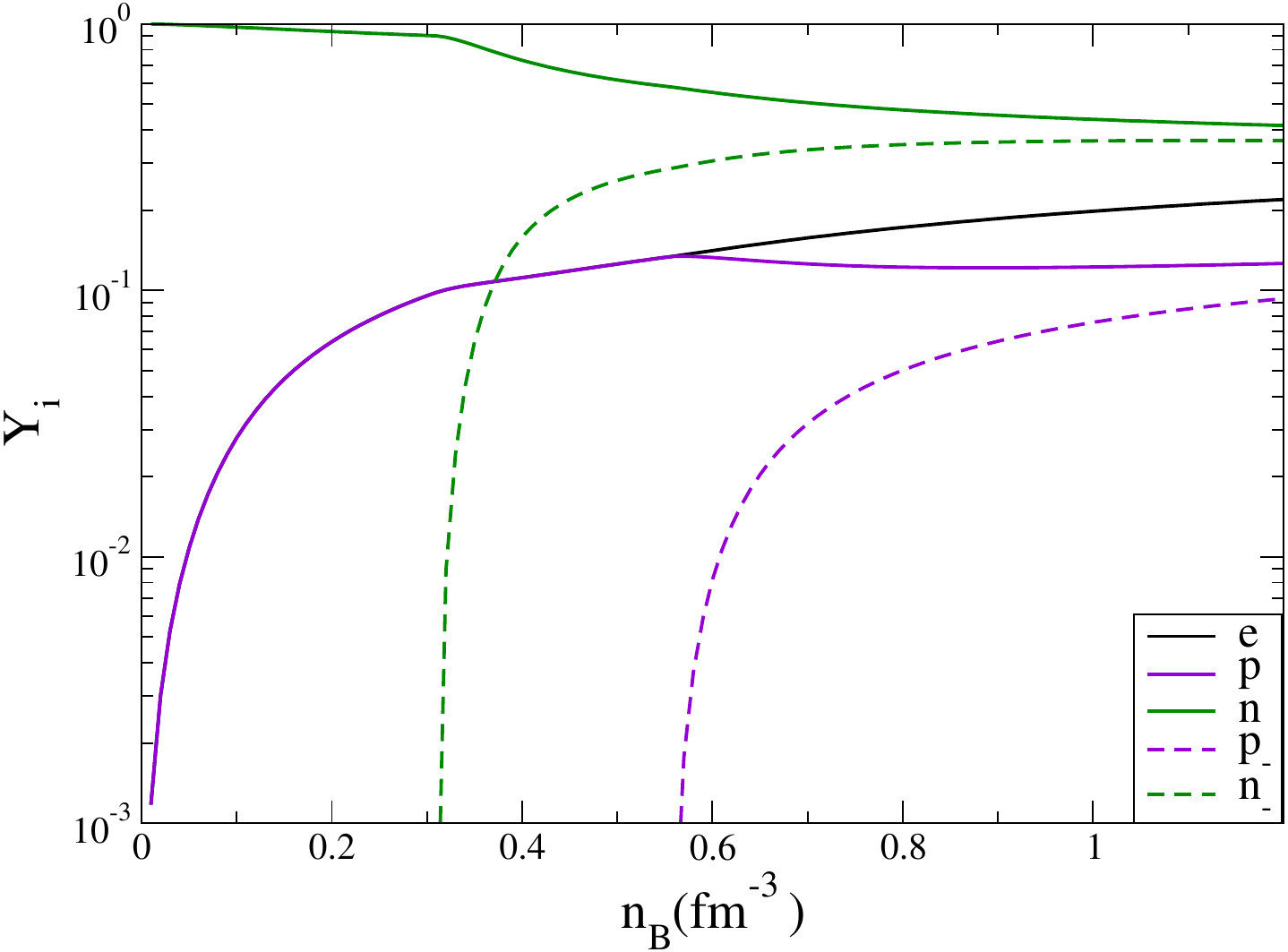}
\includegraphics[width=.396\textwidth]{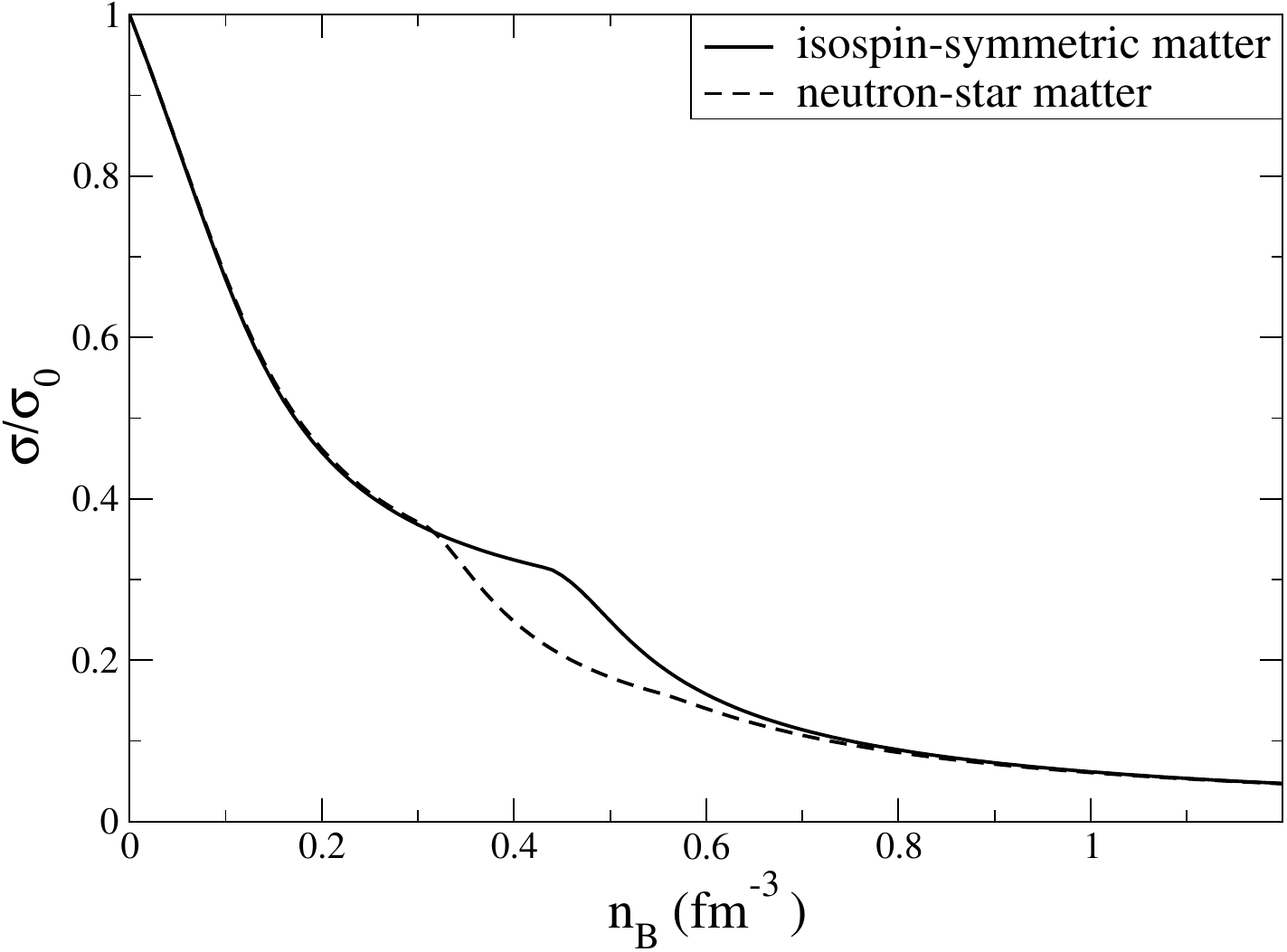}
\caption{Parity Doublet model results for particle fractions (including negative parity states) as a function of density for neutron-star matter (left) and for the chiral condensate for isospin-symmetric matter for neutron-star  matter (right). Figure adapted from~\cite{Dexheimer:2008cv}.}
\label{fig3}
\end{figure}

Chiral symmetry is a fundamental property of matter predicted by Quantum Chromodynamics (QCD), the theory that describes the fundamental building blocks of hadrons, quarks, and gluons. It is associated with the transformation that swaps left-handed and right-handed states, named in analogy to handedness - the inability to reproduce left from right hand and vice-versa using mirrored images - although, in QCD, it is defined through the relationship between particle spin and momentum. The spontaneous breaking of chiral symmetry is responsible for generating $\sim99\%$ of the mass of hadrons. At sufficiently high energies, corresponding to high $\mu_B$ or $T$ or both (and consequently high baryon density, which depends on both) chiral symmetry is expected to be approximately restored, causing the masses of particle-doublet states (hadrons with the same quantum numbers but opposite parity) to become degenerate in mass. 

Models that incorporate the spontaneous breaking and restoration of chiral symmetry are called chiral models. These models are described by Lagrangians that do not contain bare masses for the hadrons because such terms are not invariant under a chiral transformations, but instead generate hadron masses dynamically through the medium (using interaction terms that preserve chiral symmetry). Another way to say this is that the right- and left-handed components of the chiral-invariant terms transform independently. A particular class of chiral models, known as Parity Doublet models~(\cite{Detar:1988kn,Zschiesche:2006zj}), describes baryons as having large bare masses, which are only allowed because these models also include parity partners for each baryon species, e.g., a resonance with mass $M=1535~\rm{MeV}$ for the nucleons. The nucleons and their parity partners are then defined as a mixtures of two fermion fields for which the chirally-invariant masses are defined. Parity doubled models have been used to describe neutron stars for nearly 2 decades~(\cite{Dexheimer:2007tn}) and have also been extended to include parity partners of the hyperons~(\cite{Dexheimer:2012eu}). It has been shown that large bare masses can delay the appearance of hyperons in dense matter, thereby providing an alternative solution to the hyperon puzzle~(\cite{Gao:2026scv}).

The left panel of Fig.~\ref{fig3} shows the particle fractions as functions of the baryon density for charge-neutral, $\beta$-equilibrated matter in the core of neutron stars calculated within the Parity Doublet model. The parity partners of the nucleons appear at $n_B\sim2\,n_{\rm{sat}}$. The right panel shows the scalar field $\sigma$, normalized by the vacuum value, decreasing as a function of density for both neutron-star and isospin symmetric case. It is worth noting that, in chiral models, the $\sigma$ field (which generates a significant fraction of the baryon effective masses) decreases with density, signaling the restoration of chiral symmetry. For this reason, it is commonly referred to as \emph{chiral condensate}, which serves as the order parameter for chiral symmetry restoration. On the other hand, in Walecka-type models, the $\sigma$ field increases with density, although it still reduces the effective baryon masses, see Eq.~\eqref{chap2:eq13}.

In the parametrization of the Parity Doublet model adopted in~(\cite{Dexheimer:2008cv}), the phase transition is not of first order, regardless of the isospin asymmetry. In other parametrizations of the same model, the chiral-symmetry phase transition can instead be of first order, as shown in Fig.~8 of~\cite{Zschiesche:2006zj}, where the chiral condensate exhibits a ``jump'' to a much lower value with increasing $\mu_B$.
In the high $T$/low $\mu_B$ regime, first-principle lattice QCD calculations show that chiral symmetry restoration is a crossover (see Sec.~\ref{3.2}). Beyond this region, particularly in the low $T$/high $\mu_B$ regime relevant for neutron stars, the order of the chiral symmetry restoration phase transition remains unknown.

\subsection{Quark deconfinement}
\label{2.6}

In the-high $T$/low-$\mu_B$ regime, QCD calculations predict that hadrons (baryons and mesons) are deconfined into their constituent quarks. In this regime, the deconfinement phase transition is also crossover (as for the chiral symmetry restoration). Beyond this region, particularly in the low-$T$/high-$\mu_B$ regime relevant for neutron stars, the order of the deconfinement phase transition also remains unknown, although it is often speculated to become first order.

In the low-$T$/high $\mu_B$ regime, deconfinement can be easily understood following the argument of~\cite{Shapiro:1983du}. As discussed in Sec.~\ref{1.1},  assuming nucleons to have a finite hard-core volume, an individual nucleon would correspond to a number density of $n\sim0.4~\rm{fm}^{-3}$. Consequently, once the baryon density of the star is reaches this value, essentially, no empty space remains between nucleons, and beyond that density, baryons begin to overlap, making the notion of defined hadronic boundaries mostly meaningless. The central density of a neutron star increases with its mass. In the inner core of of massive neutron stars, $M\gtrsim 2~\rm{M}_{\rm{Sun}}$, depending on the EoS, central densities may reach two or three times this value (see Fig.~4 of~\cite{Tan:2021ahl} for the case of a first-order phase transition). 

\begin{figure}[t!]
\centering
\includegraphics[trim={0 0 0 0cm},width=.5\textwidth]{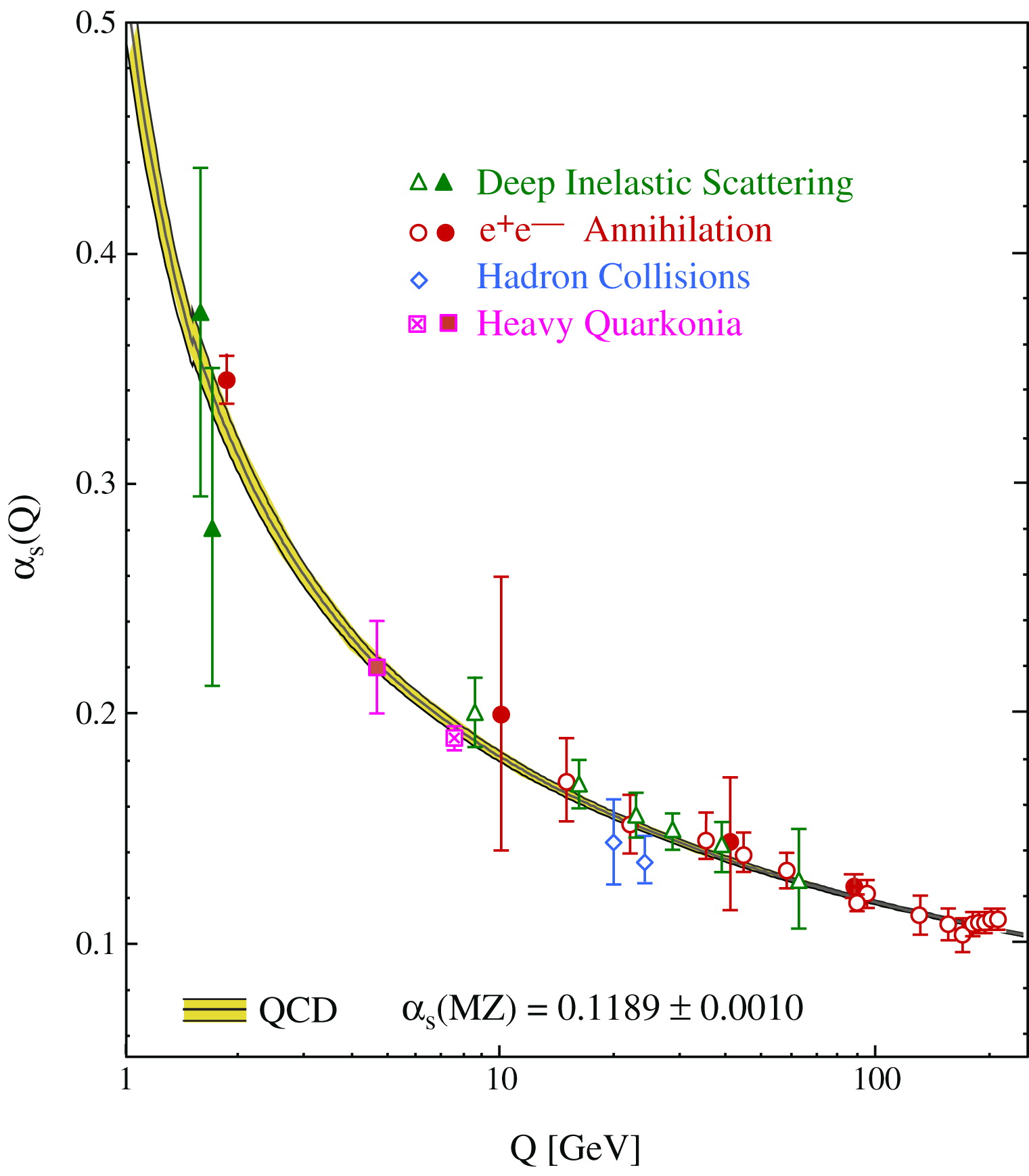}
\caption{Strength of the strong force coupling as a function of the energy scale verified by results from different experiments. Figure from \cite{Bethke:2006ac}.}
\label{fig8}
\end{figure}

A more accurate picture of quark deconfinement requires the understanding of asymptotic freedom~(\cite{Gross:1973id,Politzer:1973fx}), the most distinguished characteristic of QCD. It refers to the fact that quarks interact strongly at low and intermediate energies (including most environments in the Universe), but become weakly interacting at sufficiently high energies (extremely high-$T$ and/or high $\mu_B$ regime, well beyond the conditions expected in the core of massive neutron stars); see Fig.~\ref{fig8}. This phenomenon, known as \emph{asymptotic freedom}, arises because the carriers of the strong interaction, the gluons, themselves carry color charge, unlike photons in electromagnetism. As a consequence, gluons interact with themselves. Their self-interactions amplify the effective color charge observed at large distances, polarizing the vacuum (causing anti-screening of color charge, opposite to screening of electric charge in electromagnetism).
The traditional picture of a hadron as a bag of quarks has gradually been replaced by a deeper description of valence quarks (the traditional ones) immersed in a see of virtual quarks and gluons. Although the electric charge of the hadrons comes primarily from the valence quarks, other properties, such as their spin receive substantial contributions from the sea of quarks and gluons.~(\cite{Ji:2020ena}).
Baryons and mesons are colorless objects because they combine three colors into white, whereas mesons combine a color and a anti-color into white, leading them to behave effectively like fundamental particles at low energies.

Unfortunately, the density at which quarks are deconfined at low $T$ is still unknown, as well as what kind of phase transition it corresponds to.
In the opposite regime of high $T$ and low $\mu_B$, QCD is much better understood. This is because, in this regime, QCD can be solved numerically on supercomputers by discretizing spacetime into a lattice, with quarks forming the sites and the gluons forming the links between them. In the limit that the distance between the quarks goes to zero and the size of the lattice goes to infinity, lattice QCD (LQCD) corresponds to QCD~(\cite{Wilson:1974sk,HPQCD:2003rsu}). In this regime, quark deconfinement is known to be a crossover~\cite{Aoki:2005vt}. We return to this topic in Subsection~\ref{3.2}.

\begin{figure}[t!]
\centering
\includegraphics[width=.3\textwidth, angle=90]{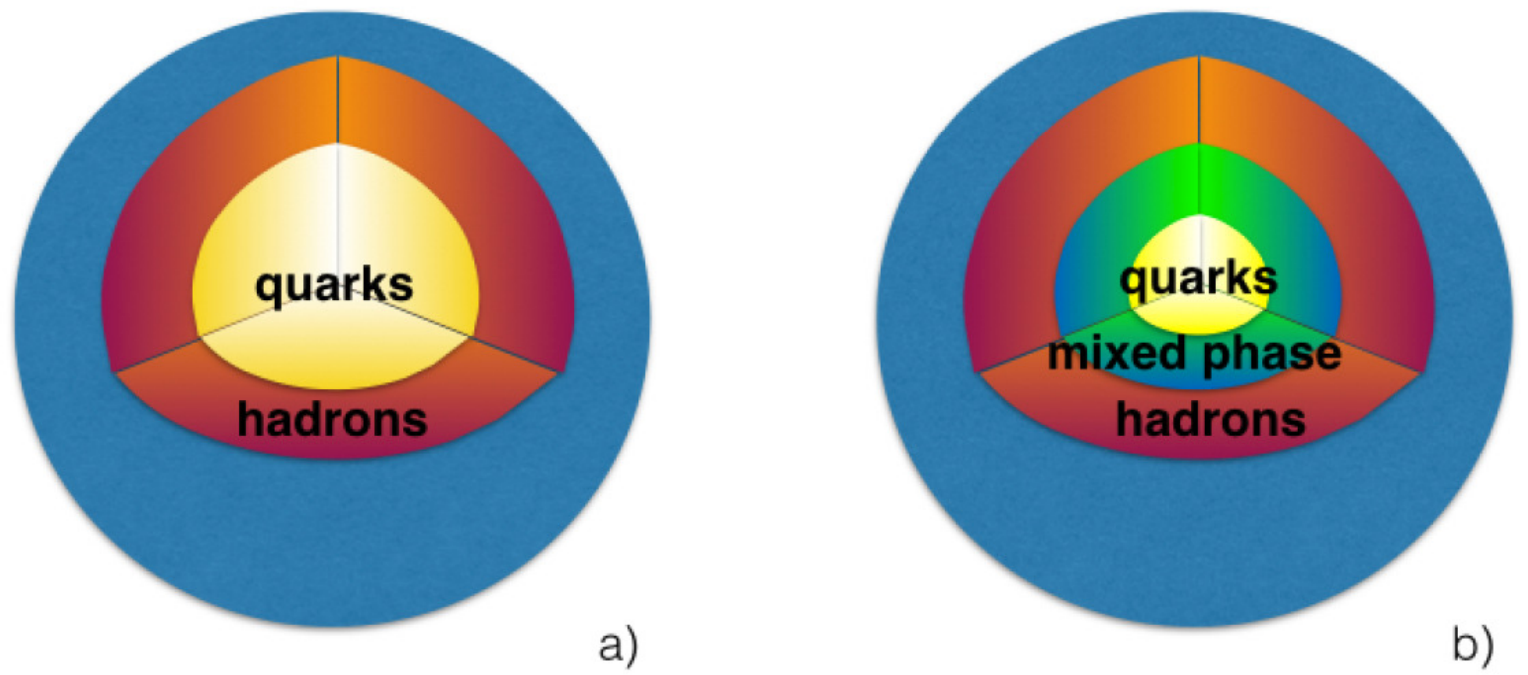}
\caption{Deconfinement to quark matter hypothesized to take place inside massive neutron stars without (left panel) and with (right panel) the presence of a mixed phase. Figure from~\cite{Contrera:2016phj}.}
\label{fig4}
\end{figure}

Back to neutron stars, deconfined quarks may appear in the inner core of massive stars, possibly giving rise to a first-order phase transition. This possibility has been discussed since the 60's~(\cite{Ivanenko:1965dg}). In this case, the interface between the hadronic and quark phases may consist either of a sharp boundary or include a mixed phase, depending on the surface tension between the phases~(\cite{Lugones:2013ema}). If the surface tension is sufficiently large, a ``wall'' will exist between the phases, whereas for sufficient low surface tension, a mixed phase with ``bubbles'' of one phase embedded in the other phase will appear. As first proposed by~\cite{Glendenning:1991hy}, such a mixed phase could occupy an extensive region within the core of neutron stars. See Fig.~\ref{fig4} for a diagram.

Within mixed phases, at least one additional degree of freedom (e.g., $\mu_Q$) must vary continuously. In neutron-star matter, each phase presents a different value of $\mu_Q$ and carries a finite (and opposite) electric charge. Only the combination of the two phases reproduce global electric-charge neutrality
\begin{align}\label{chap2:eq37}
{Y_Q}_{\rm{global}}= \lambda {Y_Q}_{\rm{Q}} + (1-\lambda) {Y_Q}_{\rm{H}}=0
\end{align}
where $\lambda$ denotes the volume fraction occupied by the quark phase, which is determined self-consistently, increasing from $0\rightarrow1$ as the density increases. Q denotes the quark phase and H the hadronic phase. Throughout the mixed phase
\begin{align}\label{chap2:eq38}
P_{\rm{H}} = P_{\rm{Q}}\,,
\end{align}
and
\begin{align}\label{chap2:eq39}
{\mu_B}_{\rm{H}} = {\mu_B}_{\rm{Q}}\,,
\end{align}
must be satisfied according to the Gibbs criteria for phase equilibrium. In this case, the hadronic phase becomes electrically positive and the quark phase negative, both being more isospin symmetric (than in the case without a mixed phase), therefore, lowering the total energy of the system.

More generally, the imposed global charge fraction may take finite values, other than ${Y_Q}_{\rm{global}}=0.5$. For ${Y_Q}_{\rm{global}}=0.5$, the isospin symmetry requires $\mu_Q=0$ (in the absence of strangeness - see~\cite{Aryal:2020ocm} for details), leaving no degree of freedom left to vary across the mixed phase. As a result, isospin-symmetric matter does not support extended mixed phases, regardless of the value of the surface tension, instead, always producing sharp interfaces between the phases. In this case, all thermodynamic derivatives of the grand potential jump (or are discontinuous) across phase boundaries. If ${Y_Q}_{\rm{global}}\neq0.5$ but the surface tension is sufficiently large, the `wall'' also implies a jump in $\mu_Q$ across the phase boundary. See~\cite{Hempel:2013tfa} and references therein for further discussions and~\cite{Roark:2018uls} for the case of mixed phases with more conserved quantities. See also Chapters 6 and 8 of~\cite{Uechi:2012wzc} for a longer discussion of phase stability in neutron stars and \cite{Lukacs:1986hu,Heinz:1987sj} for discussions in terms of strangeness for heavy-ion collisions.

A very different possibility concerns stars entirely made of deconfined quark matter, usually referred to quark stars or strange stars. These hypothetical objects first proposed by~\cite{Itoh:1970uw} are based on the hypothesis that the true ground state of matter is quark matter rather than hadronic matter. This hypothesis is commonly referred to as the Bodmer-Witten conjecture~(\cite{Bodmer:1971we,Witten:1984rs}). Although there is currently no definitive evidence for quark stars, their existence  remains a possibility. Nevertheless, if these objects exist, one must explain why ordinary matter consists of hadrons instead of deconfined quarks. Possible resolutions to this apparent paradox include the hypothesis that deconfined quark matter is truly stable only in the presence of strange quarks - whose production would require entering into contact with other strange quarks~(\cite{Farhi:1984qu}) - or, as suggested more recently, that quark matter becomes the true ground state only when a sufficient large amount of matter is assembled - as in a compact star-~(\cite{Wang:2020wzs,Wang:2025lwv}). Regardless of how quark stars would come to be, one important consequence of their existence is that, unlike neutron stars, quark stars are self-bound objects. In hadronic matter, as one goes from isospin-symmetry to asymmetry, matter becomes less bound. However, even in the isospin-symmetric matter, the maximum nuclear binding energy per nucleon is $\sim16~\rm{MeV}$, compared with $\sim200-300~\rm{MeV}$ of gravitational binding in massive neutron stars~(\cite{Glendenning:1997wn}). As a result, neutron stars are bound by gravity. Quark stars, on the other hand, are self-bound. This implies that their density does not go to zero when the pressure goes to zero (as it does in neutron stars), but instead remains finite, defining a sharp surface. Moreover, the mass-radius relation for quark stars looks quite different from that of neutron stars, with low-mass stars having small radii, whereas low-mass neutron stars typically have large radii. 

Whatever the composition of dense matter may be, both in terms of particle content and phases, it is crucial for the interpretation of  observations of massive neutron-star, which have been confidently measure to reach $\sim 2~\rm{M}_{\rm{Sun}}$ (more details in Subsection~\ref{3.4}). Therefore, it is important not only to determine the densities at which exotic hadronic matter and deconfined quark matter appear, but also the maximum density that neutron stars can reach in their cores. The difficulty is that this value depends not only on the star's macroscopic properties (such as its mass and radius), but also on the matter EoS. This problem has been addressed extensively in the literature, including discussing hybrid stars that masquerade as ordinary neutron stars~(\cite{Alford:2004pf}). 

\begin{figure}[t!]
\centering
\includegraphics[trim={2.5cm 16.7cm 0 2cm},clip,angle=270,width=.45\textwidth, angle=90]{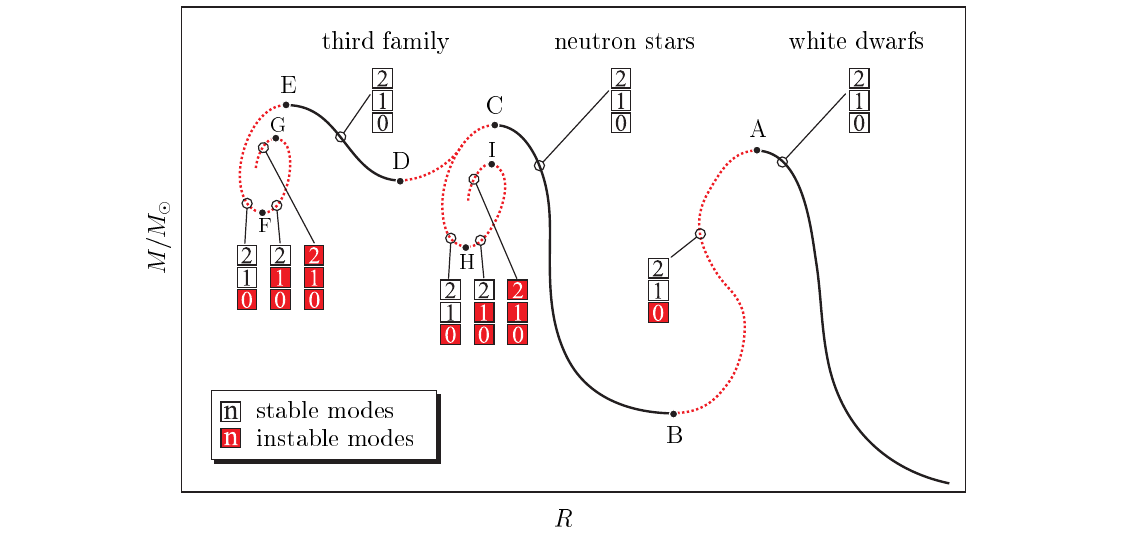}
\caption{Mass-radius diagram showing stable families of compact stars, including white dwarfs, standard neutron stars, and a possible third family of stars. Stable and unstable branches are identified. The numbers refer to the number of radial vibration modes of the star. Figure from \cite{Schertler:2000xq}.}
\label{fig45}
\end{figure}

One more possibility worth discussing is the existence of twin stars, neutron stars with the same mass but different radii.
Assuming that all neutron star cores are described by a single EoS (as opposed to, e.g, two distinct EoSs,~\cite{Drago:2015cea}), only one sequence (or line) is expected to appear in the mass-radius diagram for neutron stars. Therefore, the only possibility to create a separate stable neutron-star branch is through the presence of an intermediate unstable branch, triggered by strong phase transitions, possibly of first order (although not necessarily,~\cite{Schaffner-Bielich:2000nft,Alvarez-Castillo:2014dva}), which first softens the EoS, and is subsequently followed by a rapid stiffening. See Fig.~\ref{fig45} for a schematic mass-radius diagram showing the white-dwarf-branch, the neutron-star branch, and an additional branch, often referred to as the \emph{third branch}. Hybrid stars on the third branch that have a mass equivalent in the neutron star branch are called \emph{twin stars}. The possibility of a third branch was first discussed in books~(\cite{Harrison:1965zz,Wheeler1964Superdense}). The first paper discussing twin stars was~\cite{Gerlach:1968zz}, in both cases not in the context of deconfinement to quark matter but, rather, in the context of a phase transition from a hadronic phase to another hadronic phase with meson condensation. Much later,~\cite{Glendenning:1998ag} revived the discussion in the context of quark deconfinement. More recently, one work identified a fourth family, giving rise to triplets~(\cite{Alford:2017qgh}), triggered by two distinct first-order phase transitions: one from hadronic matter to a paired quark phase (2SC), followed by another to a different paired quark phase (CFL).

Concerning stability, it is important to note that solutions of the TOV equations guarantee only mechanical equilibrium, not stability, (analogous to a ball balanced on the top or bottom of a hill). To access stability, one must analyze the response to perturbations. In the case of stars, this means studying their oscillation modes. See~\cite{Alford:2017vca} for a concise yet thorough discussion. Summarizing, in general, a negative slope in the mass-radius diagram is associated with stability, although this criterion no longer applies beyond the maximum mass of a branch if the sequence curls inward (see the red-dotted lines in Fig.~\ref{fig45}). More refined analyzes can, of course, be performed by considering additional classes of oscillation modes, such as non-radial and gravitational modes~(\cite{Kokkotas:1999bd}). Additionally, hybrid stars may remain stable beyond the maximum mass, depending on wether the volume elements at the phase interface preserve their phase identity (slow phase transitions), or are converted from one phase to another (rapid phase transitions) as they are stretched or compressed~(\cite{Pereira:2017rmp}). 

The baryonic mass of a neutron star can be determined from the TOV equations as the number of baryons multiplied by the nucleon mass. The gravitational binding energy is then given by the difference between the (gravitational) mass and the baryonic mass, yielding  a negative value that accounts for the enormous amount of energy released during the supernovae explosions that form neutron stars.  For stars with the same baryonic mass, the configuration with the higher central density has the lower total energy. This would make the hadronic branch of a twin star configuration metastable when compared to the hybrid (or third) more stable branch. Nevertheless, stellar dynamics studies have shown that the migration between branches is possible in either direction, although is not necessarily likely. On the contrary, a star on a given branch is expected to remain on that branch unless it is subject to very large (and possibly unrealistic) perturbations~(\cite{Haque:2026ero}).

\section{Constraints on dense matter}
\label{Constraints}

\subsection{QCD Phase diagram}
\label{3.1}

\begin{figure}[t!]
\centering
\includegraphics[width=.393\textwidth,trim={0 0 -2cm 0},clip]{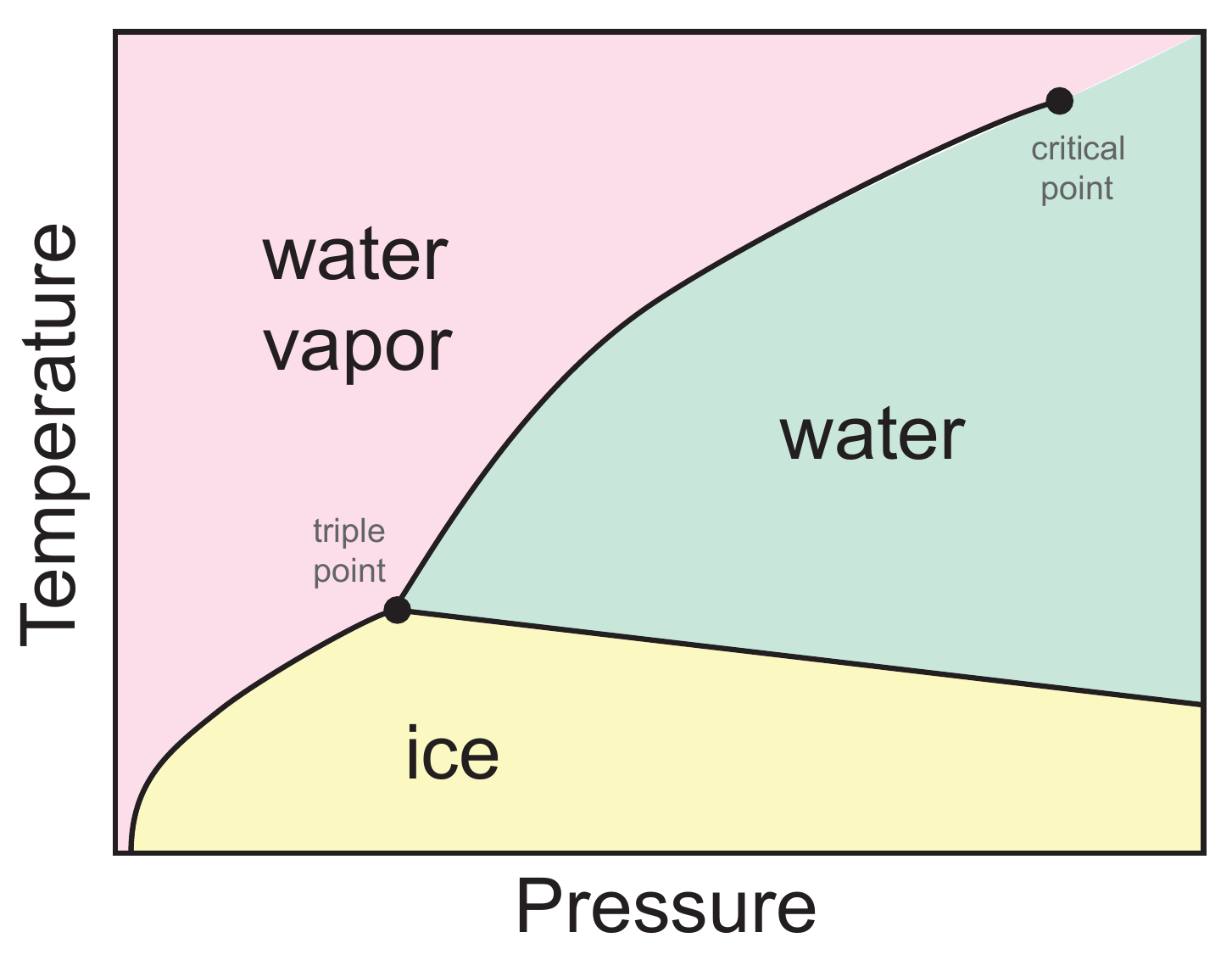}
\includegraphics[width=.414\textwidth,trim={-10cm 0 0 0},clip]{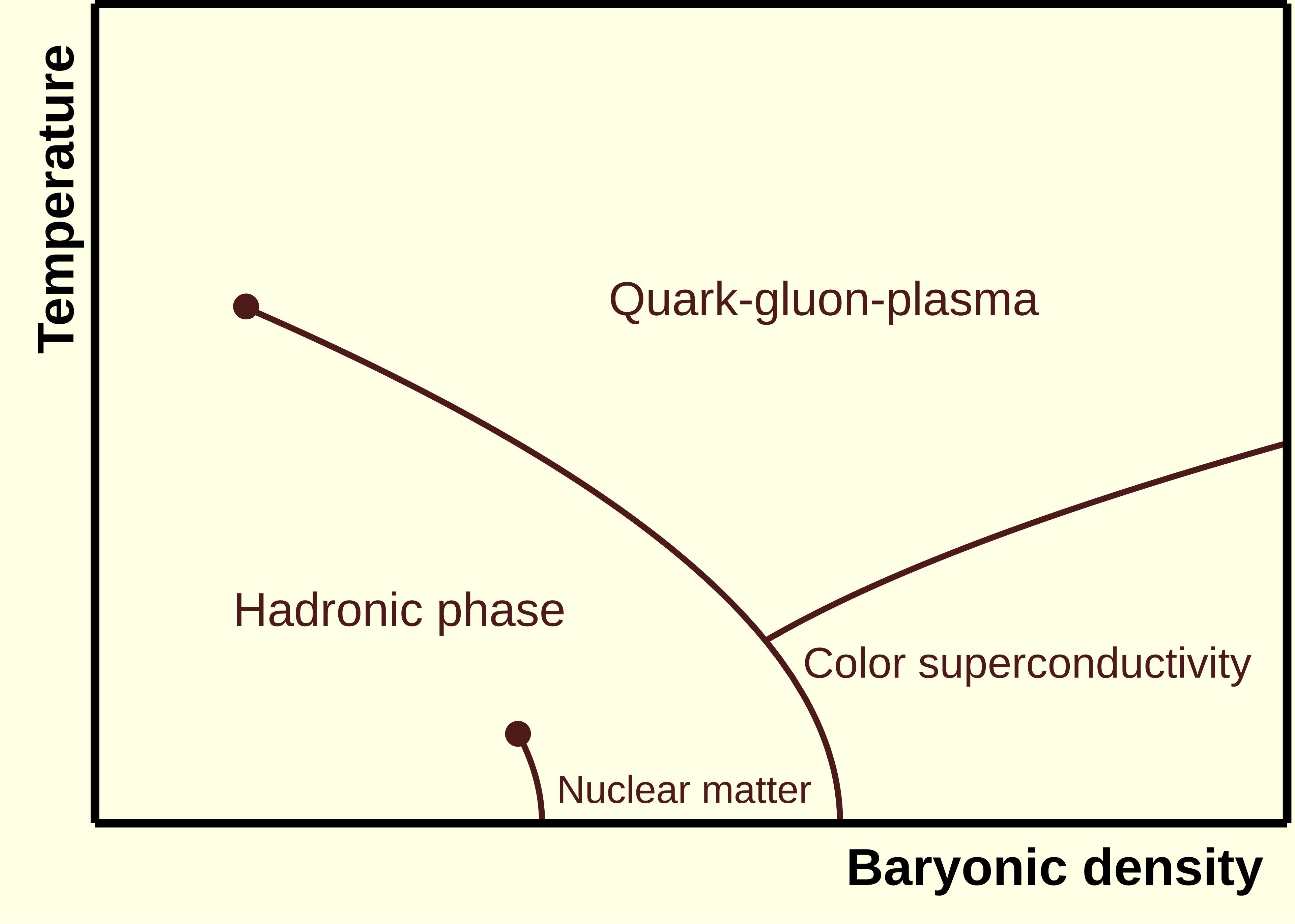}
\caption{Water phase diagram (left panel) and a sketch of the QCD phase diagram (right panel). Right panel from Wikipedia.}
\label{fig5}
\end{figure}

The easiest way to understand matter is to map all its phases onto a phase diagram. This involves choosing relevant thermodynamical quantities as the axes, identifying the different relevant phases, their boundaries, and order of the phase transitions across these boundaries (except in the case of a crossover, where no boundary exists). For example, in the well-known phase diagram for water, pressure and $T$ are chosen as the axes, and the ice, water, and water vapor phases are shown. The diagram includes first-order phase transitions (including a triple point), a second-order phase transition (the critical point), and a crossover beyond the critical point; see the left panel of Fig.~\ref{fig5}. From the water phase diagram, for a given pressure and $T$, on can immediately identify the stable phase, as well as predict what will be the stable phase as either quantity is increased or decreased.

A similar phase diagram can be constructed for strongly-interacting (high-energy) matter, the QCD phased diagram. The most relevant thermodynamical variables are $T$ and $n_B$ (often normalized by $n_{\rm{sat}}$) or $\mu_B$. The most relevant phases are nuclei, a hadronic phase (where the nuclei have dissolved into their hadronic constituents), and a quark phase (where hadrons have effectively dissolve into their constituent quarks). The boundaries between these phases include a first-order phase \emph{nuclear liquid-gas} phase transition between nuclei and hadronic matter, which becomes a crossover after a critical point, and a possible \emph{deconfinement} first-order phase transition between hadronic matter and quark matter, which may also terminate at a critical point and continue as a crossover. See the right panel of Fig.~\ref{fig5}, where quark matter at high $T$ is referred to as the \emph{quark-gluon plasma} and at low $T$ as \emph{color superconductivity} with a possible first-order phase transition separating the two. The term quark-gluon plasma steams from the fact that at high $T$ quark matter behaves as a plasma of quarks and gluons, exhibiting collective behavior~(\cite{Blaizot:2001nr}), although its transport properties more closely resemble those of a strongly coupled liquid~(\cite{Teaney:2009qa}). The term color superconductivity steams from the fact that, at high densities, quarks near the Fermi surface experience attractive interactions in certain color channels, leading to the formation of diquarks, quark Cooper pairs~(\cite{Alford:2007xm}), which may organize into several different  phases~(\cite{Warringa:2006dk}); see~\cite{Schmitt:2010pn} for a more detailed discussion. 

\begin{figure}[t!]
\centering
\includegraphics[trim={0 0 0 1.4cm},clip,width=.5\textwidth]{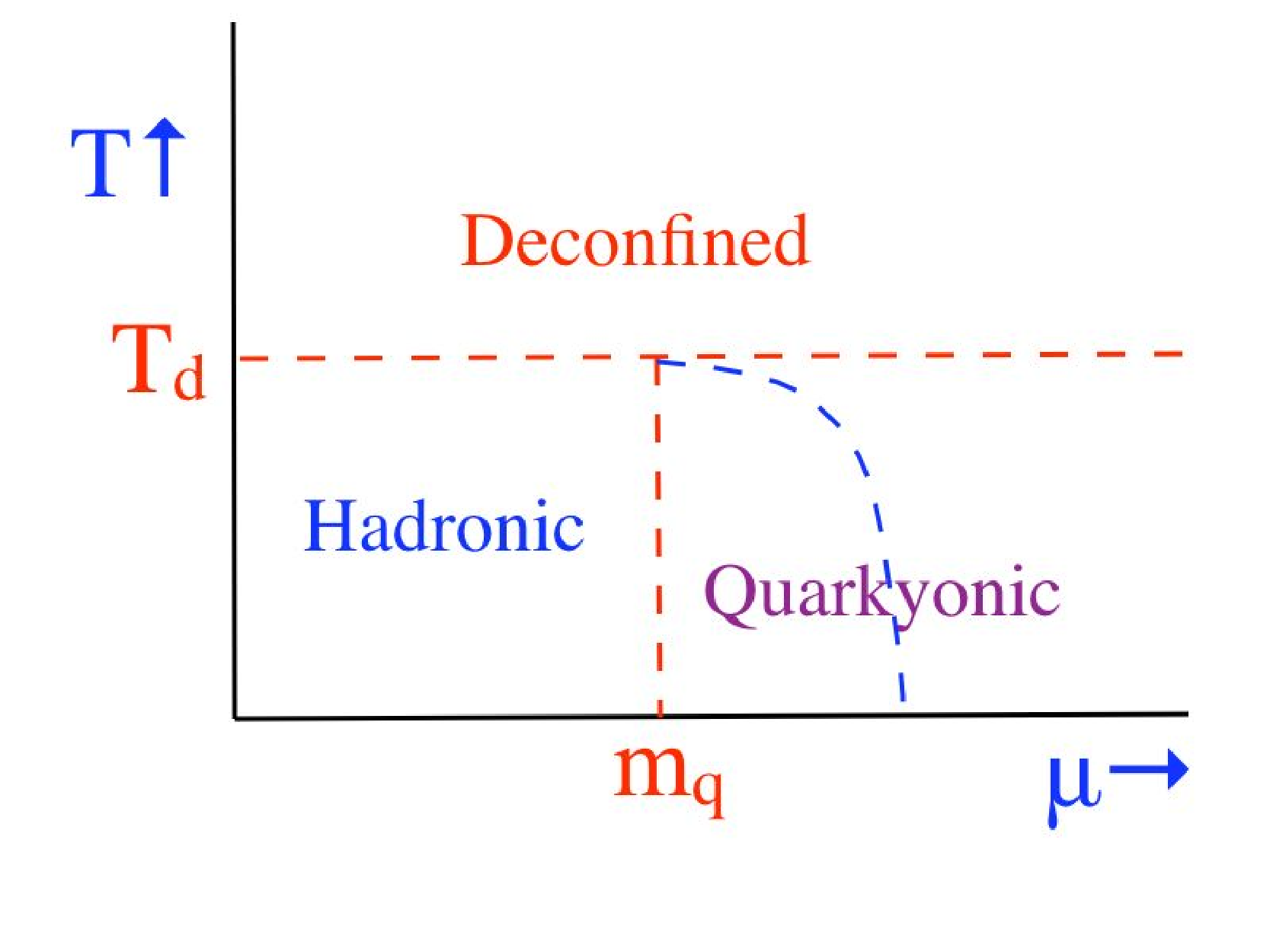}
\includegraphics[trim={0 -1.2cm 0 0cm},clip,width=.48\textwidth]{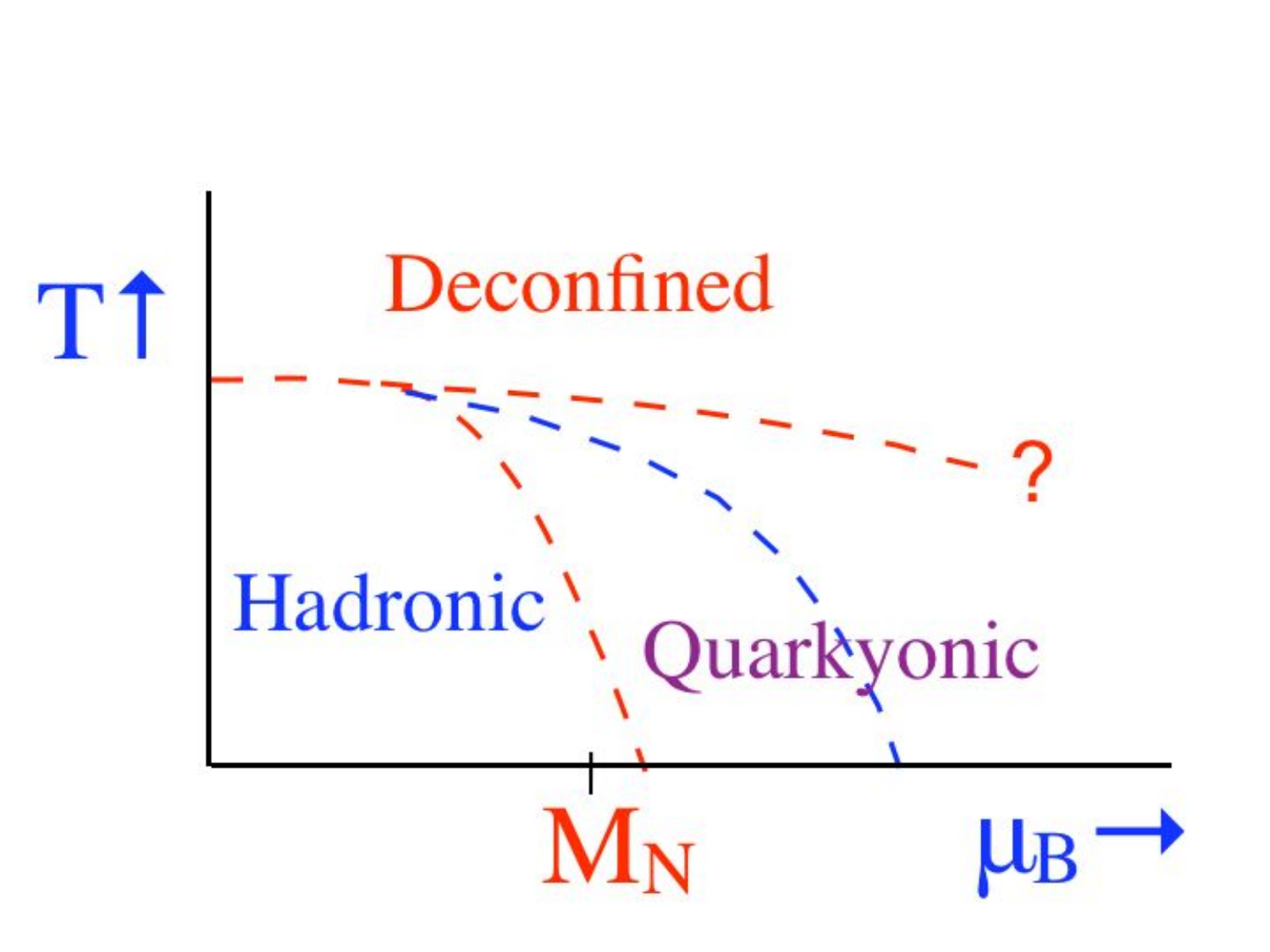}
\caption{Temperature vs. baryon chemical potential QCD phase diagrams showing the quarkyonic phase in the case of infinite colors (left panel) and 3 colors (right panel). Figures from~\cite{McLerran:2007qj}. The blue line denotes the chiral symmetry restoration and the orange line denotes quark deconfinement. $T_d$ is the deconfinement temperature, $m_q$ the quark mass, and $M_N$ the nucleon mass.}
\label{fig6}
\end{figure}

The first-order phase transition to quark matter commonly depicted in QCD phase diagrams, together with its associated critical point, are not yet experimentally confirmed. It remains possible that the transition is instead a crossover throughout the entire phase diagram. See~\cite{Baym:1979etb} for a discussion of the possibility that deconfinement occurs in stages.  In this picture, quarks first become deconfined within hadronic matter while remaining localized, followed by second transition at higher density, where they become uniformly deconfined. The mechanism is often referred to as \emph{percolation}. This idea is closely related to the concept of quarkyonic phase, in which quarks remain confined while chiral symmetry is restored~(\cite{McLerran:2007qj}); see the left panel of Fig.~\ref{fig6}. Quarkyonic matter is a prediction of QCD in the limit of a large number of colors. In our Universe, where hadrons are made up of quarks carrying $3$ colors, quarkyonic matter may still exist at low $T$ and large densities; see the right panel of Fig.~\ref{fig6}. One can understand quarkyonic matter in terms of the structure of the Fermi sea. It consists of a sea of quarks with confined (particle-hole) baryonic excitations close to the Fermi surface. As the density increases, this baryonic shell becomes progressively thinner (although the Fermi momentum increases rapidly), until it eventually disappears. Note that in this scenario there is no true phase transition. Instead, the pressure and $\mu_B$ vary rapidly, whereas the energy density and baryon density do not (in a phase transition the opposite behavior occurs).

In the following, we summarize what we \emph{do} know about the QCD phase diagram from both theory and experiment. Unfortunately, none of the available theories can currently describe the entire phase diagram, nor do the conditions explored by the different experiments, including astrophysics observations, cover it completely. As a result, large portions of the phase diagram remain unexplored, where our current knowledge relies exclusively on extrapolations or (phenomenological) models; see colored vs. white regions in Fig.~\ref{fig7}.

\subsection{From high-energy and low-energy theory}
\label{3.2}

At $\mu_B=0$, {\bf{lattice QCD}} provides a reliable determination of the EoS for $T\gtrsim 125$ MeV~(\cite{Borsanyi:2013bia,HotQCD:2014kol}).  Under such conditions, it has been established that the transition from hadronic matter (well described in this regime by a resonance gas, HRG~\cite{Hagedorn:1965st}) at low $T$ to the quark–gluon plasma at high $T$ is not a true phase transition but rather a smooth crossover~(\cite{Aoki:2005vt}). 
Although the crossover transition does not exhibit discontinuities, lattice QCD can be used to determine the $T$ at which the transition is the strongest (corresponding to the peak of the respective susceptibility), referred to as the \emph{pseudo-critical temperature}. For chiral symmetry restoration, it is $T=158.0 \pm 0.6~\rm{MeV}$~(\cite{Borsanyi:2020fev}). For quark deconfinement, the situation is more complicated because there is no exact order parameter for confinement in the presence of quarks. Nevertheless, results for the quark number susceptibilities and the Polyakov loop indicate a pseudo-critical temperature consistent, within uncertainties, with that of the chiral transition~(\cite{Bazavov:2011nk}).

\begin{figure}[t]
\centering
\includegraphics[trim={0 0 0 0cm},width=.8\textwidth]{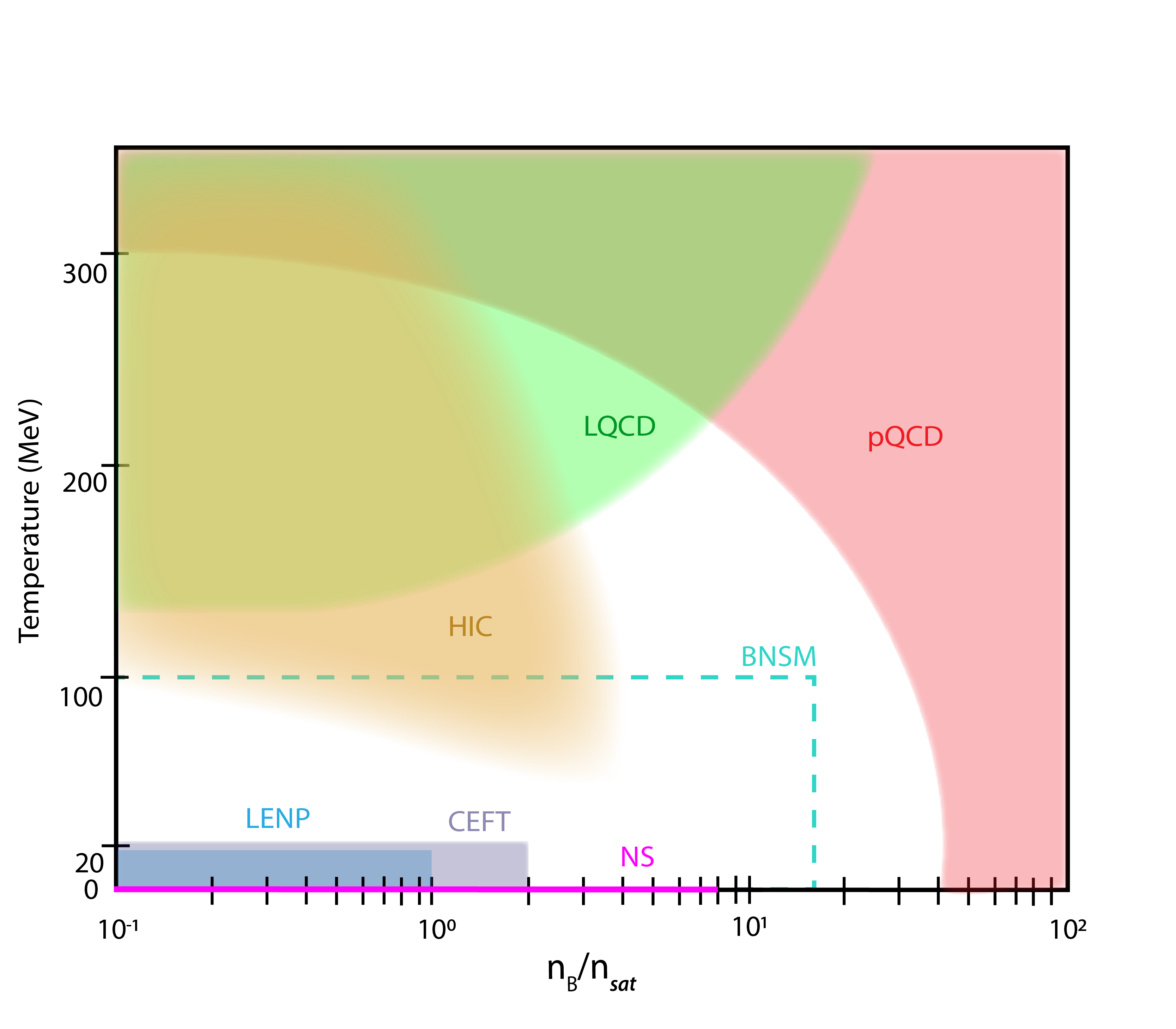}
\caption{Temperature vs. normalized baryon density QCD phase diagram showing regions where the equation of state is known from theory (lattice QCD, perturbative QCD, and Chiral Effective Field Theory) and experiments (heavy-ion collisions, low-energy nuclear physics, and neutron-star observations). Rough predictions for future detections of neutron-star gravitational-wave post-merger signals are also depicted inside dashed lines. Figure from~\cite{MUSES:2023hyz}.}
\label{fig7}
\end{figure}

At finite $\mu_B$, the QCD action acquires a complex phase, preventing its interpretation as a probability weight and rendering its numerical evaluation intractable — a difficulty known as the sign problem~(\cite{Troyer:2004ge}). Nevertheless,
one can still calculate baryon number, electric charge, and strangeness correlations using
susceptibilities (higher-orders derivatives of the pressure with respect
to the corresponding $\mu$s)~(\cite{Borsanyi:2011sw,Bellwied:2019pxh}). 
One can also perform Taylor expansions of lattice-QCD results around $\mu_B=0$, yielding a reliable EoS for values up to $\mu_B\sim 3.5\, T$~(\cite{Borsanyi:2021sxv,Borsanyi:2022qlh}). The peak in the susceptibility can also be calculated at finite $\mu_B$, allowing the determination of a pseudo-critical transition line (or simply pseudo-critical line) together with its curvature coefficients~(\cite{HotQCD:2018pds,Borsanyi:2020fev}).
Taylor-expanded lattice QCD calculations combined with the HRG model can also be used to constrain the hadronic population through the calculation of the partial pressures from different classes of particles~(\cite{Alba:2017mqu}).

To extend lattice QCD calculations to nonzero baryon chemical potential beyond expansions in $\mu_B/T$, it is possible to employ complex Langevin dynamics, originally proposed by Parisi~\cite{Parisi:1983mgm}. In this approach, complex configurations are evolved in a fictitious time (rather than considering many configurations) and sampled via a stochastic process; see e.g. the lecture notes by~\cite{Aarts:2015tyj} and chapter in this book by~\cite{Aarts:2026uiu}. Recently, first results for the QCD equation of state have been obtained above the crossover temperature ($T \geq 260 MeV$) and for unprecedentedly high baryon densities ($\mu_B/T \lesssim 10$)~(\cite{Mandl:2026ngb}).

Finally, although lattice QCD has not yet identified a critical point associated with either the chiral symmetry phase transition or quark deconfinement, one can use this null result to exclude a critical point up to $\mu_B \approx 300~\rm{MeV}$~(\cite{HotQCD:2019xnw}). On the other hand, owing to the negative curvature of the pseudo-critical line around $\mu_B=0$, the $T$ of the critical point (if it exists) is expected to be lower than the pseudo-critical temperature at $\mu_B=0$. In the $2+1$ chiral limit case (zero light quark masses but non-zero strange quark mass), this implies that the crossover could become a second-order phase transition at a larger $\mu_B$ only for $T < 132^{+3}_{-6}~\rm{MeV}$~(\cite{HotQCD:2019xnw}).
Although the lattice-QCD expansion in $\mu_B$ can cover a large portion of the QCD phase diagram when shown in terms of $n_B$ (which also depends on $T$), it cannot reach the lower right side of Fig.~\ref{fig7}. Nevertheless, it can still be used to constrain quark couplings in neutron-star models, as it was first demonstrated in 2009 within the CMF model~(\cite{Dexheimer:2009hi}).

At asymptotically high energies, QCD predicts that the strong interaction becomes weakly coupled; see again Fig.~\ref{fig8}. As a consequence of asymptotic freedom, at sufficiently large $T$ and/or $\mu$s, perturbation theory becomes applicable to QCD
and the corresponding analytic calculations (known as {\bf{Perturbative QCD}} or pQCD) become reliable~(\cite{Haque:2014rua,Ghiglieri:2020dpq}). Where applicable, pQCD provides not only the EoS but also many other properties, as discussed below.

In practice, applying perturbation theory to QCD requires resummations to all orders of the perturbative expansion. The two
main methods used to accomplish such resummations are effective field theory~(\cite{Braaten:1995cm,Braaten:1995jr}) and hard-thermal-loop perturbation theory~(\cite{Andersen:2002ey,Andersen:2003zk}). In particular, the hard-thermal-loop technique accounts for the fact that, at high $T$, the plasma strongly modifies particle propagation. 
A loop integral is the momentum-space integral associated with a closed loop in a Feynman diagram. It represents the contribution of virtual quarks and gluons to a physical process. Since the momentum flowing around the loop is not fixed by momentum conservation for virtual particles, one must integrate over all possible loop momenta. The problem is that the low momentum part of the loop integrals generates infrared divergences associated with long-range color fields that should be screened. The solution to these divergencies is to resum separately the dominant thermal corrections arising from these diagrams, known as hard thermal loops.

Both resummation schemes
have been extended to next-to-next-to leading order (N2LO) in the perturbation expansion at $\mu_B=0$. At finite $\mu_B$, N2LO~(\cite{Freedman:1976xs,Freedman:1976dm,Freedman:1976ub}) and partial next-to-next-to-next-to-leading order (N3LO) results are also available at $T=0$~(\cite{Gorda:2018gpy,Gorda:2021znl,Gorda:2021kme}).
As a result, pQCD becomes applicable at $T \gtrsim 300~\rm{MeV}$ at $\mu_B = 0$ and at $n_B \gtrsim 40~n_{\rm{sat}}$ at
$T = 0$. Between these limits, pQCD has been extended to different combinations of $\mu_B$ and $T$ across the phase diagram, always at sufficiently high energies~(\cite{Kurkela:2016was}). Although such large densities (at low $T$) are well beyond those found in neutron-star cores, it has been shown that causality and stability bounds allow
pQCD to constrain matter at lower densities~(\cite{Komoltsev:2021jzg}). But, most importantly,  thermodynamic results from pQCD agree with lattice QCD data for $T \gtrsim 2-3~T_c$~(\cite{Andersen:2011sf}). PQCD can also predict the curvature of the pseudo-critical phase transition line~(\cite{Haque:2020eyj}), in agreement with lattice QCD.
Going beyond thermodynamic equilibrium quantities, pQCD has been used to compute the bulk viscosity (measuring how far the system departs from equilibrium when its volume changes)~(\cite{Arnold:2006fz}), shear viscosity (the same, but associated with fluid layers sliding past one another)~(\cite{Ghiglieri:2018dib,Danhoni:2022xmt}), relaxation time (how soon the system can be well described by hydrodynamics)~(\cite{Ghiglieri:2018dgf}), conductivity (how quickly electric charge is conducted),  diffusion (how conserved charges spread)~(\cite{Arnold:2000dr}), and second-order transport coefficients (which determine how quickly a system returns to equilibrium after being perturbed)~(\cite{York:2008rr}) were also computed.

\begin{figure}[t!]
\centering
\includegraphics[trim={0 0 0 0cm},angle=270,width=.5\textwidth]{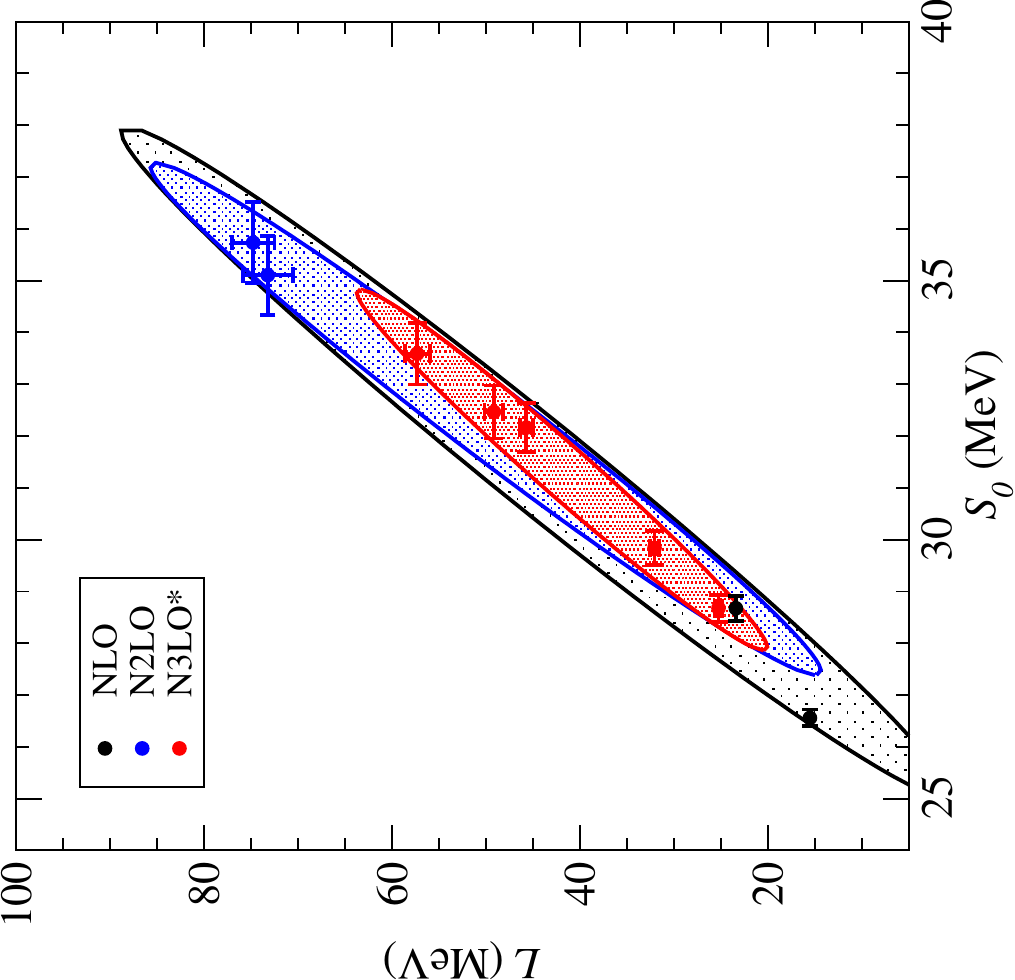}
\caption{Correlation between the symmetry energy (here called $S_0$) and its slope at saturation from CEFT calculations performed at different orders in the chiral expansion. Figure from~\cite{Holt:2016pjb}.}
\label{fig9}
\end{figure}

At low energies, quarks are no longer the relevant degrees of freedom. In this regime, {\bf{Chiral effective field theory}} ($\chi$EFT or CEFT)
offers a systematic and model-independent framework for describing hadronic matter with quantifiable uncertainties~(\cite{Weinberg:1978kz}). The theory starts from writing  the most general Lagrangian consistent with spontaneously broken chiral symmetry, using nucleons and pions as the relevant degrees of freedom. One then performs an order-by-order expansion of the interactions in momentum. While the long-range features of the theory are governed by pion-exchange contributions constrained by chiral symmetry, the short-distance details are encoded in a set of contact interactions whose strengths are fitted to scattering and bound-state experimental data. Recently, advances in the application of Bayesian statistical methods
have enabled robust uncertainty quantification in EoS calculations up to fourth order in the chiral expansion, applicable to densities up to $n_B\lesssim 2~n_{\rm{sat}}$~(\cite{Hebeler:2010jx,Sammarruca:2014zia,Tews:2018kmu,Drischler:2020hwi}). Beyond that, one can extrapolate to higher densities using, e.g., the many-body non-perturbative Brueckner-Hartree-Fock approach, which resums repeated two-body scattering processes~(\cite{Logoteta:2016hxh,Logoteta:2016nzc}). 

At $n_{\rm{sat}}$, CEFT can be used to determine the symmetry energy
\begin{align}\label{chap3:eq40}
E_{\rm{sym},\rm{sat}}=\left(\frac{E_{Y_Q=0}-E_{Y_Q=0.5}}{A}\right)_{n_{\rm{sat}}}\,,
\end{align}
which is the difference in energy per nucleon between pure neutron matter and isospin-symmetric nuclear matter at saturation. Its slope parameter is defined as
\begin{align}\label{chap3:eq41}
L_{\rm{sat}}=3n_{\rm{sat}}\left(\frac{dE_{\rm{sym}}}{dn_B}\right)_{n_{\rm{sat}}}\,.
\end{align}
The results are shown in Fig.~\ref{fig9} and quantify how much stiffer isospin-asymmetric matter becomes as a function of density, providing an essential constraint for modeling the interior of neutron stars.

Chiral effective field theory has also been extended to finite $T$ ($T<25~\rm{MeV}$) and employed to study the nuclear liquid-gas phase transition in isospin-symmetric nuclear matter~(\cite{Wellenhofer:2014hya}), yielding critical values $T_c\sim17-19~\rm{MeV}$ with corresponding densities $n_c\sim0.6-0.8~/\rm{fm}^3$.

Note that, from the theoretical perspective, the results discussed in this Subsection can be extrapolated to different isospin asymmetries and strangeness contents. In the case of pQCD, as long as the energy is sufficiently high, there is no distinction between large $\mu_B$, $\mu_Q$, and $\mu_S$, since the calculations are performed in terms of $\mu$s for the quarks, which combine the $\mu_B$, $\mu_Q$, and $\mu_S$. Lattice QCD is a somewhat different, as the sign problem persists for $\mu_S$, but not for $\mu_Q$. For the latter, recent calculations cover the entire $T$-$\mu_I$ phase diagram predicting several different phases, including pion condensation~(\cite{Son:2000xc}). Note that one can define a formulation in which the isospin chemical potential is equal to the char one, $\mu_I=\mu_Q$~(\cite{Aryal:2020ocm}). CEFT works similarly to pQCD, in the sense that, as long as the energy is low, the calculations can be extended to finite $\mu_Q$. Including strangeness in CEFT calculations, on the other hand, is more complicated because hyperons tend to appear at low $T$ around $n_B=2~n_{\rm{sat}}$, where CEFT stops being applicable; see~\cite{Haidenbauer:2019boi} for a discussion of hyperon-nucleon interactions within CEFT. The CEFT EoS for neutron stars can also be used to inform other neutron-star EoSs (e.g., by fitting couplings,~\cite{Dexheimer:2018dhb}) or produce new parametric or non-parametric EoSs, which can subsequently be connected to pQCD~(\cite{Kurkela:2014vha}).

\subsection{From high-energy and low-energy experiment}
\label{3.3}

Starting from the high-energy side, energetic {\bf{heavy-ion collisions}} provide information over a wide range of temperatures $T\sim 50-650~\rm{MeV}$ and densities, depending on the stage of the collision and the center-of-mass beam energy of the collision $\sqrt{s_{NN}}$. Higher collision energies probe higher $T$s and lower densities, whereas lower collision energies probe lower $T$s and higher densities. The anti-correlation with density arises because, in more energetic collisions (referred to as relativistic because the nuclei travel at velocities close to the speed of light), the nuclei pass right through one another. In this case, the matter produced after the collision contains nearly equal numbers of particles and antiparticles ($\mu_B\sim0$). In less energetic collisions, the nuclei are said to \emph{stop}, in which case the original baryon density and isospin asymmetry of the original nuclei become relevant.

\begin{figure}[t!]
\centering
\includegraphics[trim={0cm 1cm 0cm 12cm},clip,width=.6\textwidth]{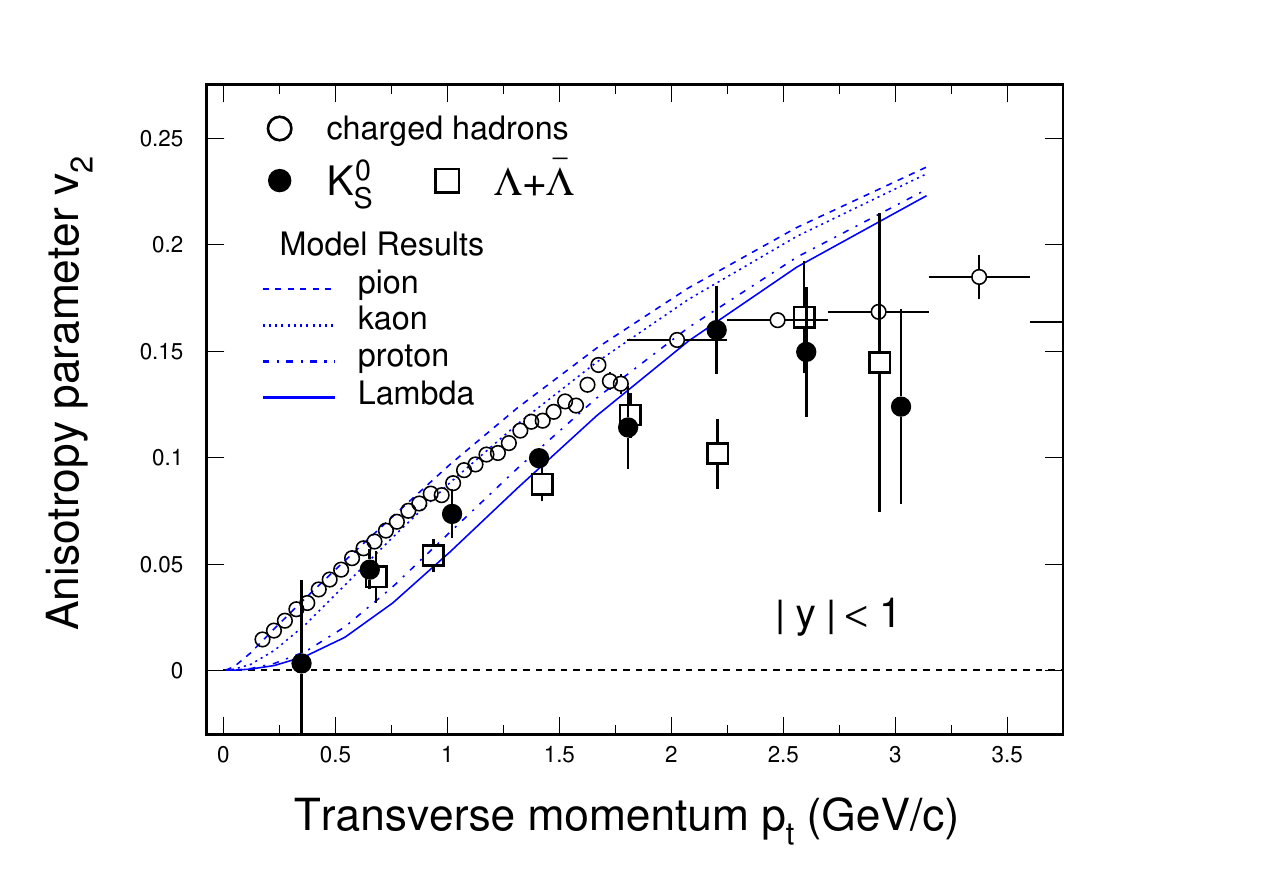}
\includegraphics[trim={0 0cm 0 1cm},clip,width=.8\textwidth]{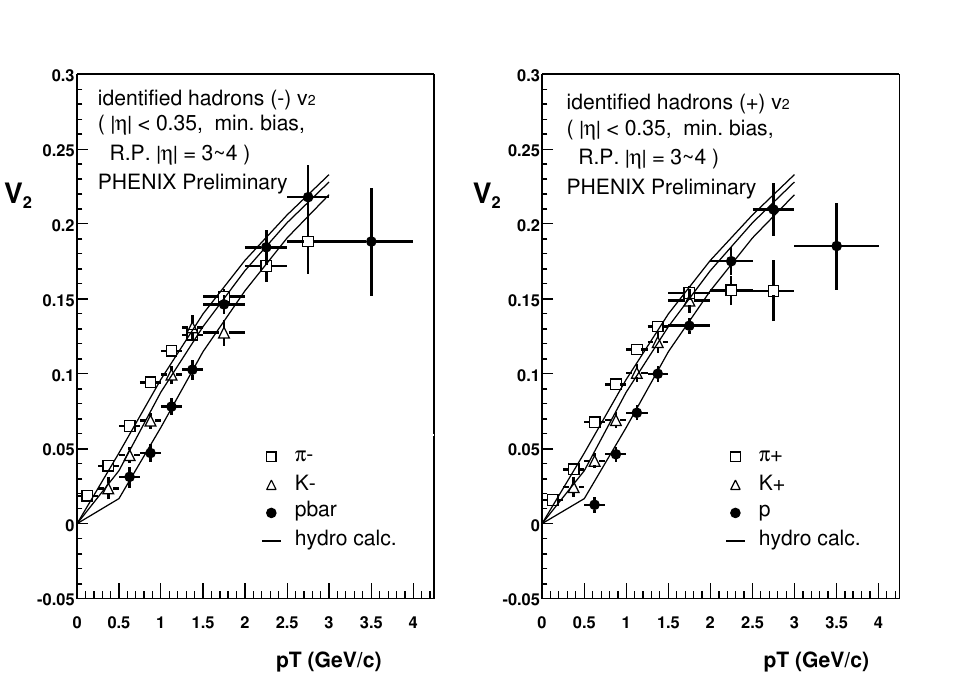}
\caption{Elliptic flow coefficient $v_2$ data as a function of transverse momentum obtained from heavy-ion collisions. Hydrodynamical predictions are also shown. Figures from the Star collaboration (top,~\cite{STAR:2002okv}) and PHENIX collaboration  (bottom,~\cite{Esumi:2002vy}).}
\label{fig10}
\end{figure}

Although the particles detected experimentally are produced during the later stages in the collision (after the \emph{chemical freeze-out}, when particle abundances no longer change), they can be used to infer the properties of earlier stages.  Collision $T$s and $\mu$s can be extracted using, e.g., thermal models, which provide a statistical-equilibrium description in which the hadronic population depends only on $T$ and $\mu$s~\cite{Wheaton:2004qb}). The emission of photons and lepton pairs (dileptons)~(\cite{Gale:2014dfa}) can also probe $T$, since they do not interact via the strong force and can therefore traverse the medium largely unmodified. Heavy-quark bound states (containing charm and/or bottom quarks) can also probe $T$, particularly during the early stages of the collision~(\cite{Matsui:1986dk}). 
At the highest collisions energies, deconfined quark matter can be produced, referred to in this context as the quark-gluon plasma. Evidence for this comes, e.g., from strangeness production, where the creation of strange quark anti-quark pairs is much easier at the quark level than in hadronic interactions~(\cite{STAR:2007cqw,ALICE:2013xmt}), although this is observed experimentally later, through the enhanced production of strange mesons and strange baryons.

However, the strongest evidence for the formation of the quark-gluon plasma comes from the analysis of (collective) flow. Flow refers to the collective motion of  particles, which do not move independently in the hot and dense medium. This behavior originates from the initial almond-shaped collision geometry (resulting from the overlap of the colliding nuclei), which generates pressure gradients within the system. The initial spatial anisotropy is rapidly ($\sim10^{-24}~{\rm{s}}$) converted into an anisotropy in momentum space, which governs the subsequent evolution of the system. The particle distribution is commonly expanded in a Fourier series, with coefficients $v_1$, $v_2$, $v_3$, etc corresponding to directed flow, elliptic flow, triangular flow, and higher-order harmonics. 

In particular, $v_2$ played an essential role in establishing evidence for the formation of the quark gluon plasma~(\cite{PHENIX:2004vcz, PHOBOS:2004zne, STAR:2005gfr}). Because the initial spatial anisotropy disappears rapidly as the system expands, $v_2$ is generated predominantly during the earliest stages of the collision~(\cite{Ollitrault:1992bk}). Results from both PHENIX, PHOBOS, and STAR at the Relativistic Heavy Ion Collider (RHIC at Brookhaven National Laboratory), the first relativistic heavy-ion collider built, point to the formation of a strongly interacting, nearly perfect fluid (with extremely low viscosity) at the earliest stages of the collisions, inconsistent with hadronic models. See Fig.~\ref{fig10}, where mesons and baryons exhibit the same dependency on momentum, indicating that dynamics are governed by the underlying quark degrees of freedom. The curves for baryons and mesons tend to overlap when scaled by the number of quark constituents; see Fig.~20 of~\cite{STAR:2005gfr}. Fig.~\ref{fig10} also shows that hydrodynamics provides an excellent description of the system up to high momenta, beyond which it gradually breaks down.

Further evidence for the production of the quark-gluon plasma, as well as additional information about dense and hot matter come from jet quenching (where high-energy quarks and gluons lose energy while traversing the medium,~\cite{Gyulassy:1990ye}), Hanbury Brown–Twiss (HBT) interferometry (where correlations between particle pairs provide a way to map the system~,\cite{Heinz:1999rw, Wiedemann:1999qn, Lisa:2005dd}), quarkonium suppression (where heavy-quark bound states melt because of color screening,~\cite{Digal:2001ue,Braun-Munzinger:2000csl}), and fluctuations. 
Fluctuations are sensitive to critical phenomena because they scale strongly with the correlation length, which measures the distance over which different parts of a system influence one another. This is particularly important in the search for a critical point associated with quark deconfinement and chiral-symmetry restoration phase transitions (see, e.g.,~\cite{Stephanov:2008qz,Fraga:2010qef,Mroczek:2020rpm}), where the susceptibilities are expected to diverge and exhibit nontrivial behavior in its vicinity. Additionally, quantities associated with fluctuations, such as cumulants of particle distributions (statistical measures of the event-by-event distributions of particle species or conserved quantities) can be related to thermodynamic susceptibilities calculated in lattice QCD~(\cite{Parotto:2018pwx,Bazavov:2007zz}). 

Note that all the experimental observables discussed above can be used, to varying degrees, to constrain the EoS of hot and dense matter. In the following, we focus on on lower energies, including heavy-ion collisions as well as other types of nuclear experiments that do not break apart nuclei. Before continuing, it is worth mentioning that collective flow in low-energy heavy-ion collisions can also provide important constraints for the neutron-star EoS~(\cite{Danielewicz:2002pu,Spieles:2020zaa,Sorensen:2022odd}).

{\bf{Low-energy nuclear physics}} provides a wealth of information around $n_{\rm{sat}}$ for isospin-symmetric matter and beyond. 
The most important quantity we discuss is $n_{\rm{sat}}$ itself, the density at which all the other parameters are discussed. It can be extracted, e.g., by dividing the number of nucleons in a nucleus by its volume~(\cite{1981A&A...102..299H}), yielding  
$n_{\rm{sat}}=0.17\pm0.03~\rm{fm}^{-3}$. The quoted uncertainty comes exclusively from the uncertainty in the measurement of the nuclear radius measurements from electron scattering and muonic-atom experiments (where replacing an electron with a $\mu$ brings the orbit much closer to the nucleus)~(\cite{Myers1977}). 
In a relativistic approach, one can alternatively use the relation between baryon density, Fermi momentum, and binding energy per nucleon to obtain $n_{\rm{sat}}=0.148-0.185~\rm{fm}^{-3}$,
as inferred from the Coester line~(\cite{Gross-Boelting:1998qhi}). Finally, parity-violating asymmetry experiments can also be used to constrain $n_{\rm{sat}}$. From measurements of the elastic scattering of
longitudinally polarized electrons from $^{208}$Pb,  the Lead Radius Experiment (PREX) collaboration reported $n_{\rm{sat}}=0.1480\pm0.0038~\rm{fm}^{-3}$~(\cite{PREX:2021umo}), where the quoted uncertainty includes both theoretical and experimental contributions.

In nuclear physics, the binding energy per nucleon is defined as the energy required to separate a nucleus into its constituents. Starting from the semi-empirical mass formula
\begin{align}\label{chap3:eq42}
    B=a_v A-a_s A^{2 / 3}-a_c \frac{Z(Z-1)}{A^{1 / 3}}-a_a \frac{(N-Z)^2}{A}+\delta(A, Z)\, ,
\end{align}
where the first term represents volume effects, the second term surface effects, the third term accounts for Coulomb interactions, the fourth term describes the effect of isospin asymmetry, and the last term accounts for pairing effects~(\cite{Martin:2019}). Of particular importance is the coefficient of the first term, $a_{\mathrm{v}}$, since for infinite isospin-symmetric nuclear matter, the binding energy per nucleon becomes $B/A \sim a_{v}$. 
The value $B/A$=-15.677 MeV at $n_{\rm{sat}}=0.16146~\rm{fm}^{-3}$ was obtained from a four-parameter mass formula with coefficients from the liquid-drop model using experimental masses of 49 heavy nuclei~(\cite{Myers:1966zz}). Later, a value  of $B/A$=-16.24 MeV at $n_{\rm{sat}}=0.16114~\rm{fm}^{-3}$ was obtained from a four-parameter mass formula with coefficients from shell-corrected Thomas-Fermi model using experimental data from 1654 ground state masses of nuclei with $N, Z\ge 8$~(\cite{Myers:1995wx}). 

\begin{figure}[t!]
\centering
\includegraphics[trim={0 0 0 0cm},clip,width=\textwidth]{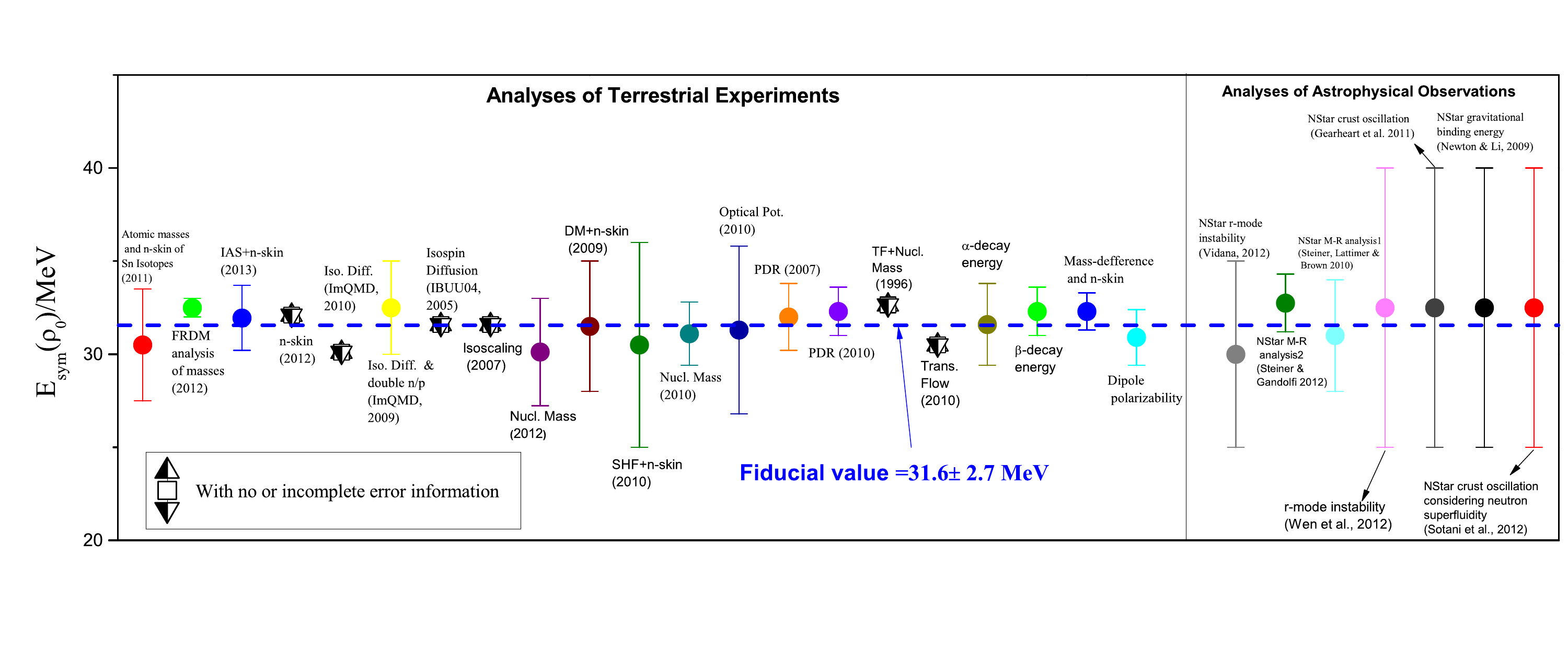}
\includegraphics[trim={0 0 0 0cm},clip,width=\textwidth]{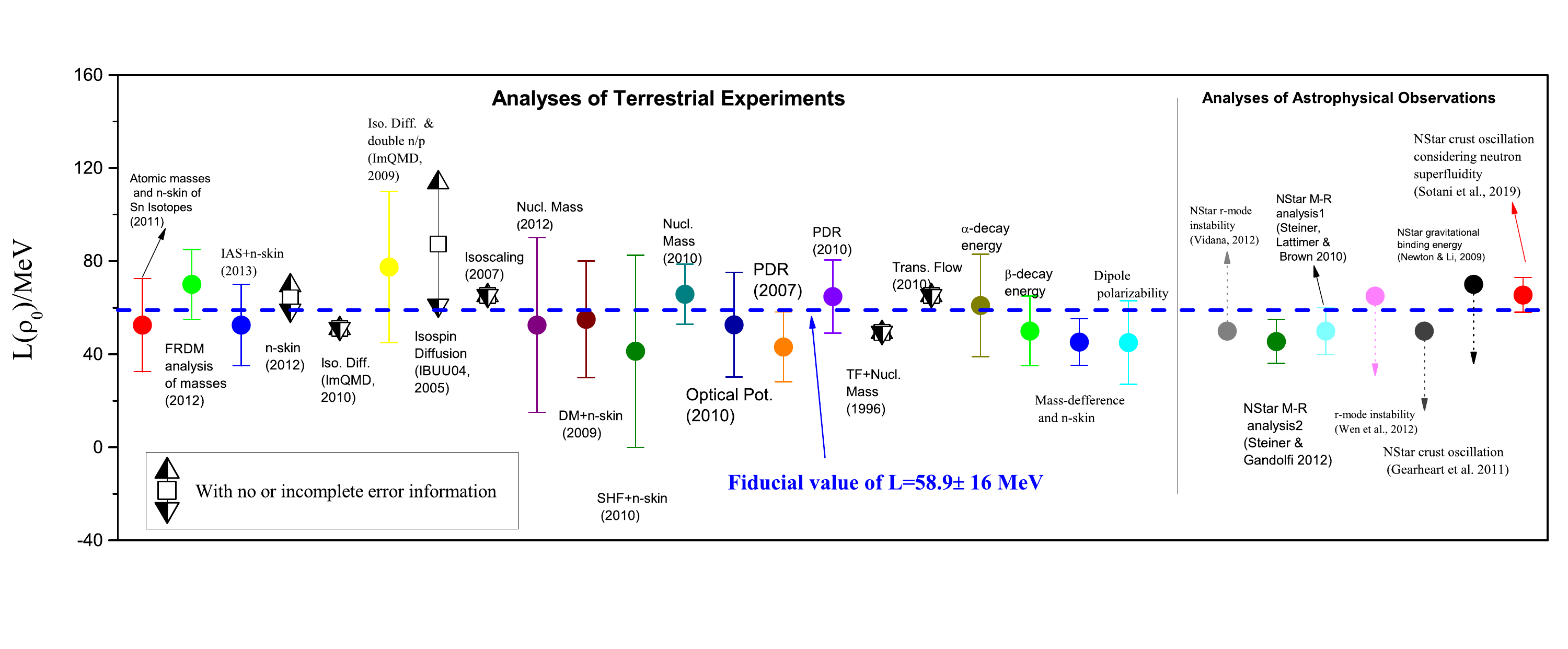}
\caption{Compilation of symmetry energy constraints (top) and slope parameter (bottom). Figures from \cite{Li:2019xxz}. Horizontal blue lines show fiducial values and analyzes involving astrophysics are shown to the right.}
\label{fig11}
\end{figure}

The compressibility (see Eq.~\eqref{chap2:eq16}) at $n_{\rm{sat}}$, also referred to at saturation as the compression modulus, can be expressed in terms of the liquid-drop mass formula
\begin{align}\label{chap3:eq43}
K_{A}=K_{\mathrm{vol}}+K_{\mathrm{surf}} A^{-1 / 3}+K_{\mathrm{asym}} \Big(\frac{N-Z}{A}\Big)^{2}+K_{\mathrm{Coul}} \frac{Z^{2}}{A^{4 / 3}}\, ,
\end{align}
where the first term is associated with volume effects, the second with surface effects, the third with isospin asymmetry, and the fourth with Coulomb effects. As for the binding energy, for infinite systems $K\sim K_{\mathrm{vol}}$~(\cite{Blaizot:1980tw}). In finite nuclei,  the isoscalar giant monopole resonance is directly related to the compressibility~(\cite{Colo:2013yta}). Experiments involving $^{90}$Zr and $^{208}$Pb nuclei suggested a value $K=240\pm20$ MeV~(\cite{Colo:2013yta,Todd-Rutel:2005yzo,Colo:2004mj,Agrawal:2003xb}), noting that pairing effects might have softened this value~(\cite{Cao:2012dt,Vesely:2012dw}).
Conversely, \cite{Khan:2012ps} argued that nuclear properties constrain the EoS not at $n_{\rm{sat}}$ but rather at an average density $\sim0.11~\rm{fm}^{-3}$, which they termed the crossing density. At this density, they obtained $K = 230 \pm 40~\rm{MeV}$ from measurements of giant monopole resonance in a chain of nuclei, corresponding to an uncertainty of $\sim17\%$ uncertainty.
A separate study carried out a comprehensive review of compressibility values extracted using different theoretical approaches and experimental techniques between 1961 and 2016~(\cite{Stone:2014wza}). It reported values spanning the range $K=100-380~\rm{MeV}$, with relativistic mean-field models generally favoring larger values than non-relativistic ones. Using an approach largely independent of microscopic model assumptions, they obtained $250 < K_{\infty} < 315~\rm{MeV}$ and demonstrated that surface properties play a key role.

Going beyond nucleons, the experimental properties of heavier baryons, including hyperons and $\Delta$-baryons, have also been measured. The optical potentials, $U$, of hyperons and $\Delta$ baryons in isospin-symmetric matter at saturation provide information about the balance between the attractive and repulsive components of the strong interaction for these particles. While experiments indicate that the $\Lambda$ potential is attractive (with $U(\Lambda-N)\equiv U_\Lambda=-28~\rm{MeV}$ from~\cite{Millener:1988hp} and $U_\Lambda=-30~\rm{MeV}$ from~\cite{Gal:2016boi}), there are indications that the $\Sigma$ potential is repulsive~(\cite{Gal:2016boi}), whereas the $\Xi$ potential appears to be  attractive ($U_\Xi=-14~\rm{MeV}$ from~\cite{Gal:2016boi} and $U_\Xi = 21.9\pm 0.7~\rm{MeV}$ from~\cite{Friedman:2021rhu}). Furthermore, the $\Delta$ potential has been inferred to lie within the range $-90~\rm{MeV}~<U_{\Delta}<-50~\rm{MeV}$~(\cite{Drago:2014oja}, combining data from~\cite{Horikawa:1980cv,Koch:1985qz}).

Going beyond symmetric matter, while still at saturation, measurements related to the symmetry energy (Eq.~\eqref{chap3:eq40}) can be used to calibrate dense-matter models. While many independent measurements of the symmetry energy cluster around a value of $32~\rm{MeV}$, the corresponding values of its slope, Eq.~\eqref{chap3:eq41}, exhibit a much wider spread. See Fig.~\ref{fig11} for a compilation of experimental and theoretical results for both quantities~(\cite{Li:2019xxz}), with fiducial values of $E_{\rm{sym}}=31.6\pm2.7~\rm{MeV}$ and $L=58.9\pm16~\rm{MeV}$. Particularly interesting are measurements of the neutron-skin thickness (the region near the surface of neutron-rich nuclei where there is an excess of neutrons). Different analyses of the PREX-II results for $^{208}$Pb reported either a large slope, $L=106\pm37~\rm{MeV}$~(\cite{Reed:2021nqk}), or a smaller value, $L=54\pm8~\rm{MeV}$~(\cite{Reinhard:2021utv}), while CREX reported a similarly small value for $^{48}$Ca, $ L=53\pm13~\rm{MeV}$~(\cite{CREX:2022kgg}). Note that measurements of both the symmetry energy and its slope are also available below $n_{\rm{sat}}$;  see~\cite{Sorensen:2023zkk} for details. Finally, there is a correlation between the symmetry energy and its slope~(\cite{Lattimer:2014sga}), which can be useful when using these quantities to constrain dense-matter models.

At finite $T$, nuclear experiments provide information on the coexistence line and critical point of the nuclear liquid-gas phase transition in nearly isospin-symmetric matter. As $T$ increases, the surface tension of nuclei decreases, eventually vanishing~(\cite{Landaustatbook}). The critical $T$ is found to lie in the range $T_c=15-23~\rm{MeV}$~(\cite{Karnaukhov:2003vp,Elliott:2013pna,Natowitz:2002nw,Karnaukhov:2008be}). While this phase transition is expected to be a strong first order in symmetric matter, in neutron stars the combined effects of isospin asymmetry and Coulomb interactions can modify the picture significantly, giving rise to mixed phases (see~\cite{Hempel:2013tfa} and references therein), including the formation of nuclear pasta phases~(\cite{Ravenhall:1983uh}). 

\subsection{From observation}
\label{3.4}

\begin{table}[t!]
\caption{Neutron-star mass and radius observations from NICER analyzed by two different groups.}
\label{mass-radius}
\centering
\begin{tabular}{ccccc}
\hline 
Object & $M(\rm{M}_{{\rm{Sun}}})$ & Radius (km) & Group & Reference   \\ 
\hline
PSR J0030+0451
 & $1.43^{+0.20}_{-0.17}$&  $12.68^{+1.31}_{-1.04}$ & Amsterdam & \cite{Kini:2026rjx}   \\ 
PSR J0030+0451
 & $1.44^{+0.15}_{-0.14}$&  $13.02^{+1.24}_{-1.06}$ & Illinois-Maryland & \cite{Miller:2019cac}  \\ 
PSR J0740+6620
 & $2.073^{+0.069}_{-0.069}$&  $12.49^{+1.28}_{-0.88}$  & Amsterdam & \cite{Salmi:2024aum}  \\ 
PSR J0740+6620 & $2.08^{+0.07}_{-0.07}$&  $12.92^{+2.09}_{-1.13f}$ & Illinois-Maryland & \cite{Dittmann:2024mbo}   \\
PSR J0437+4715 & $1.418^{+0.037}_{-0.037}$&  $11.36^{+0.95}_{-0.63}$ & Amsterdam & \cite{Choudhury:2024xbk}   \\   
PSR J0437+4715 & $1.418^{+0.044}_{-0.044}$&  $11.8-15.1$ & Illinois-Maryland & \cite{Miller:2025qfq}   \\    
\hline
\end{tabular}
\end{table}

Although the EoS is itself an equilibrium property, it can be directly related to neutron star properties. For proto-neutron stars and neutron star mergers, this relation is more challenging, but still possible. First, effects associated with departures from chemical equilibrium can be implemented in a straightforward way at the EOS level (which becomes 2-dimensional), at least for microscopic EoSs. Even when the EoS is not microscopic (e.g., a parametric EoS), effects beyond chemical equilibrium can be incorporated using expansions~(\cite{Yao:2023yda}), including the cases of finite strangeness~(\cite{Yang:2025wop}) and finite $T$~(\cite{Mroczek:2024sfp}). When finite $T$ is included in the EoS, thermal equilibrium is usually treated in astrophysics by considering each stellar layer separately and assuming that each layer is in local thermal equilibrium. At the EoS level, one either assumes some sort of $T$ profile (e.g., fixed entropy per baryon~(\cite{Prakash:1996xs}) which does not increase the dimensionality) or provide a full 2-dimensional table (3-dimensional if matter is out of chemical equilibrium). This correspond, e.g., to \emph{the general tables} provided by the CompOSE EoS repository for astrophysics~(\cite{Oertel:2016bki,Typel:2013rza}); see the instruction manuals in~\cite{CompOSECoreTeam:2022ddl,Dexheimer:2022qhn}.

Mechanical equilibrium, or the lack thereof, is treated at the macroscopic level. In the case of neutron stars and proto-neutron stars, the TOV equations~(\cite{Tolman:1939jz,Oppenheimer:1939ne}) enforce mechanical equilibrium, whereas supernova explosions and neutron-star mergers require specialized hydrodynamical simulations that account for general relativity. These codes can also include contributions beyond the ideal-gas approximation, shear and bulk viscosities; see, e.g.,~\cite{Most:2022yhe,Chabanov:2023blf,Zappa:2022rpd}.

Although neutron stars emit electromagnetic radiation across the spectrum, their emission typically peaks in the X-ray band, which can be used to probe several of their properties, including their masses. The most accurate neutron-star masses measurements come from timing  radio pulsars in binary systems that exhibit relativistic effects, most notably \emph{Shapiro delay}, an additional time delay experienced by a light signal (or radio pulse) as it propagates through curved spacetime surrounding a massive object. The first such measurement for a very massive star was PSR~J1614-2230, initially measured to have a mass of $\sim2~\rm{M}_{\rm{Sun}}$, but later refined to $1.937\pm 0.014~\rm{M}_{\rm{Sun}}$ at the $1\sigma$ ($68.3\%$) credibility level~(\cite{NANOGrav:2023hde}).
Currently, the most massive neutron star with a precise mass measurement is PSR~J0740+6620, with $M=2.08_{+0.07}^{-0.07}~\rm{M}_{{\rm{Sun}}}$, reported at the $68.3\%$ credibility level~(\cite{Fonseca:2021wxt}). Finally, the third most massive neutron star with a precise mass measurement is PSR~J0348+0432. It was previously reported to have a mass of $2.01\pm 0.04~\rm{M}_{\rm{Sun}}$, but then later reanalyzed, yielding a revised mass of $1.806\pm 0.037~\rm{M}_{\rm{Sun}}$  at the $68.3\%$ credibility level~(\cite{Saffer:2024tlb}).
Although more massive neutron stars have been reported (e.g., black-widow pulsars), these results may be affected by systematic uncertainties that are not yet quantified. Other objects may turn out to be black holes. We return to the latter possibility in Section~\ref{conclusions}. 

The most reliable neutron-star radius measurements come from NASA's Neutron Star Interior Composition Explorer (NICER) through X-ray pulse-profile modeling, namely the analysis of X-ray pulses produced by hot spots on the star's surface using general-relativistic effects (primarily light bending) that depend on the star's compactness, $M/R$. See Tab.~\ref{mass-radius} for a summary of the NICER results reported by two independent analysis groups. Note that these uncertainties correspond to $68\%$ confidence regions, which are not rectangular, but instead have more complicated shapes because of correlation between the fitted parameters. Some entries in Tab.~\ref{mass-radius} also incorporate information from XMM-Newton, the Parkes Pulsar Timing Array (PPTA), and the Nuclear Spectroscopic Telescope Array (NuSTAR).

Note that neutron-star masses and radii are also constrained by causality. Mass-radius sequences cannot extend beyond the limit (towards larger masses and smaller radii) defined by an EoS with $c_s^2=1$. This is discussed in detail in \cite{Schaffner-Bielich:2020psc}, together with the Buchdahl limit (which defines a maximum compactness), and the black-hole limit (corresponding to the event horizon of a black-hole).

Concerning neutron star mergers, the LIGO-Virgo-KAGRA (LVK) collaboration has so far confirmed $6$ relevant detections; see Tab.~\ref{mergers} for details, including at least $2$ binary neutron-star mergers and $3$ neutron star-black hole mergers. Notably, the first detection, GW170817, was triply special, as it was the only event confirmed to be accompanied by electromagnetic counterparts~(\cite{LIGOScientific:2017ync}), and the only one for which the tidal deformability (the distortion induced by the gravitational field of a close companion) has been measured. The electromagnetic counterparts observed across the spectrum included the \emph{kilonova}~(\cite{Metzger:2019zeh}), a transient event roughly $1,000$ times brighter than a classical nova, yet still much dimmer than a supernova. The kilonova emission is powered by the radioactive decay of neutron-rich ejected ($\sim0.05~\rm{M}_{{\rm{Sun}}}$), allowing the properties of the ejecta to be characterized. In particular, kilonovae are highly efficient sites for the production of heavy elements beyond iron~(\cite{Kasen:2017sxr}). Finally, the implications of the kilonova for the fate of the GW170817 merger remnant (which eventually collapsed to a black hole) could constrain the EoS through inferences of the neutron-star radius and maximum mass~(\cite{Bauswein:2017vtn,Margalit:2017dij,Radice:2017lry,Rezzolla:2017aly,Ruiz:2017due,Shibata:2017xdx}).

\begin{table}[t!]
\caption{Neutron-star mergers detected by the LVK collaboration in the different runs, including double neutron-star mergers and neutron star-black hole mergers, showing the primary and secondary object masses.}
\label{mergers}
\centering
\renewcommand{\arraystretch}{1.15}
\begin{tabular}{cccccccc}
\hline
Run & Dates                & BNS & NS-BH & Events & Prim. $M(\rm{M}_{{\rm{Sun}}})$ & Sec. $M(\rm{M}_{{\rm{Sun}}})$ & Reference\\ 
\hline
O1  & Sep 2015 -- Jan 2016 & -- & -- & -- & -- & --  & --\\
O2  & Nov 2016 -- Aug 2017 & 1 & -- & GW170817 & $1.46^{+0.12}_{-0.10}$ & $1.27^{+0.09}_{-0.09}$ & \cite{LIGOScientific:2017vwq}\\
O3a & Apr 2019 -- Oct 2019 & 1 & -- & GW190425 & $1.60-1.87$ & $1.46-1.69$ & \cite{LIGOScientific:2020aai}\\
O3b & Nov 2019 -- Mar 2020 & -- & 3 & GW191219 & $31.1^{+2.2}_{-2.8}$ & $1.17^{+0.07}_{-0.06}$ & \cite{KAGRA:2021vkt}\\
    &                      &  &  &  GW200105 & $8.9^{+1.2}_{-1.5}$   & $1.9^{+0.3}_{-0.2}$ & \cite{LIGOScientific:2021qlt}\\
    &                      &  &  &  GW200115 & $5.7^{+1.8}_{-2.1}$   & $1.5^{+0.7}_{-0.3}$ &  \cite{LIGOScientific:2021qlt}\\
O4a & May 2023 -- Jan 2024 & 1 \rm{or} & 1 & GW230529 & $3.6^{+0.8}_{-1.2}$   & $1.4^{+0.6}_{-0.2}$& \cite{LIGOScientific:2024elc}\\
O4b & Apr 2024 -- Jan 2025 & -- & -- & -- & -- & -- & --\\
\hline
\end{tabular}
\end{table}

No neutron-star merger measurement has so far extended into the post-merger phase~(\cite{LIGOScientific:2017fdd}) because of higher frequencies of the gravitational waves, expected to lie in the kHz range, and their even smaller amplitudes compared with the inspiral signal, making it  more difficult to detect. Nevertheless, the LVK collaboration was able to extract very important information from the inspiral waveform of GW170817. This included the chirp mass (a nontrivial combination of the individual masses; see~\cite{ColemanMiller:2021lky} for its derivation) and the binary tidal deformability, $\tilde{\Lambda}$, which depends on the component spins. Assuming that the neutron stars had low spins yielded $\tilde{\Lambda} = 300^{+500}_{-190}$, whereas relaxing this assumption resulted in $\tilde{\Lambda} \leq 630$, both at the $90\%$ confidence level~(\cite{LIGOScientific:2018hze}).

To determine the individual stellar masses and tidal deformabilities, one would need additional information about the merger (not yet available with current detector sensitivity) or employ a statistical approach~(\cite{LIGOScientific:2018hze,Essick:2019ldf}).  Alternatively, one can use the approximate EoS-insensitive, so-called \emph{universal relations} (e.g., the I-Love-Q relation,~\cite{Yagi:2013awa}, and binary Love relation,~\cite{Xie:2022brn}) for this purpose~(\cite{Chatziioannou:2018vzf}). It is important to note that several universal and quasi-universal relations have been proposed, with varying degrees of accuracy; see, e.g., \cite{Tan:2021nat} for a discussion.
Furthermore, there is a relation between tidal deformability and the radius (larger stars are more deformable,~\cite{Raithel:2018ncd}), but this relation is not universal and depends, to some extent, on the details of the EoS~(\cite{Dexheimer:2018dhb}). Nevertheless, one can still attempt to map the tidal deformability to neutron-star radius using universal relations~(\cite{LIGOScientific:2018cki,De:2018uhw}) or methods for constructing model-independent EoSs (e.g., spectral or non-parametric EoSs, or Gaussian processes,~\cite{Yagi:2016bkt}).

Other neutron-star properties can be used to constrain dense matter, e.g., those related to thermal evolution (or cooling) of neutron stars. Stars in which the Urca process is active cool much more rapidly~(\cite{Yakovlev:2004iq,Page:2004fy}), with the Urca process threshold itself depending on the EoS~(\cite{Lopes:2024bvz}) and on the degrees of freedom present at the relevant density~(\cite{Prakash:1992zng}). This degeneracy, which only becomes more severe when pairing is added to the list of unknowns, makes it difficult to extract definitive EoS constraints directly from cooling.

\section{Conclusions and outlook}
\label{conclusions}

Over the last century, humanity has gone from believing that the atom was indivisible to understanding its quantum orbits and nuclear structure, measuring the properties of individual hadrons, and identifying their constituents, the quarks and the gluons. The effort to understand how quarks and gluons interact to form hadrons is supported by numerous laboratories that operate particle colliders, which are used to study elementary particles and interactions and to test theoretical models against experimental data. This effort is challenged by the fact that the deconfined state produced in such experiments has a lifetime of less than one sextillionth of a second.
At lower collision energies, the collisions last longer and probe matter that is colder and farther from isospin symmetry. Nevertheless, laboratory experiments will not, in the foreseeable future, reproduce the conditions found inside neutron stars, namely cold, ultra-dense matter. This is where neutron-star observations become essential, since observing neutron stars with different masses allows us to probe different densities, extending all to several times nuclear saturation density. The main challenge is to disentangle the observational information in order to extract the EoS of dense matter, since many different EoSs can reproduce the same neutron-star mass and radius~(\cite{Clevinger:2025acg}).

In recent years, the nuclear astrophysics community has moved from describing the dense-matter EoS as simply \emph{overall soft} or \emph{overall stiff} to recognizing that EoSs exhibit much richer structure. These features are are more clearly seen in the derivative of the EoS, $c_s^2$. As suggested more than one decade ago in \cite{Chamel:2012ea,Alford:2013aca} and later discussed in detail by \cite{Bedaque:2014sqa}, $c_s^2$ of cold dense matter does not only exhibit non-monotonic behavior, but is also expected to exceed the value $\sqrt{1/3}$ before decreasing again. At asymptotically high $T$ and/or $\mu_B$, matter behaves as an ideal quantum gas of massless, non-interacting particles and the EoS simply becomes $P=\varepsilon/3$, with $c_s^2=\sqrt{1/3}$, which is referred to in the literature as the \emph{conformal limit} or the \emph{Stefan–Boltzmann limit}. Structure in $c_s^2$, often referred to as \emph{bumps}~(\cite{Tan:2020ics}), naturally satisfies the requirements for the neutron-star EoS to be soft at low densities (as implied by the small radii and tidal deformabilities of intermediate-mass neutron-stars), stiff at high densities (to support massive neutron stars), and to return asymptotically to $\sqrt{1/3}$ without ever violating the causal limit ($c_s^2=1$). However, the structure of $c_s^2$ is still far from being fully understood. See, e.g.,~\cite{Mendes:2026kpt} for a recent discussion on the need of two bumps and \cite{Hippert:2024hum} for a new limit of $c_s^2=0.781$ that was recently proposed as an empirical bound for relativistic systems in which the relevant transport coefficients can be computed. 

In recent years, driven by our new ability to detect gravitational waves emitted by neutron-star and black-hole mergers, a complementary area of research has emerged, focused on understanding the dense matter created during these events. So far, this has provided a new constraint on dense matter through the measurement of the tidal deformability during the inspiral. In the near future, we expect that new new detectors within the LVK collaboration, as well as next-generation interferometers, will be able to measure gravitational waves from the post-merger phase of the signal (in the kHz range), which can directly probe the dense-matter $c_s^2$~(\cite{Hammond:2025kki}). Even more importantly, the post-merger gravitational-wave signal is sensitive to finite $T$ effects, reaching $T$s comparable to those achieved in heavy-ion collisions~(\cite{HADES:2019auv}), probing departures from weak equilibrium~(\cite{Alford:2017rxf,Alford:2019kdw,Alford:2019qtm,Alford:2021lpp,Gavassino:2020kwo,Celora:2022nbp,Most:2022yhe}),  as well as  the deconfinement transition to quark matter~(\cite{Most:2018eaw,Bauswein:2018bma,Constantinou:2021hba}).
In addition, space-based gravitational-wave detectors will be able to observe deci-Hz frequencies, detecting merger events months or even years before coalescence and alerting telescopes in advance to observe the resulting kilonovae~(\cite{Berti:2026riu}). Lower-frequency gravitation-wave detectors from LISA~(\cite{LISA:2022yao}) will also be able to detect white-dwarf mergers, which may (unlike neutron-star and black hole mergers) produce stable neutron stars, opening an entire new avenue of research.

Until then, the large unexplored regions between the available first-principle theories, experiments, and observations (shown in the QCD phase diagram in Fig.~\ref{fig7}) can be described by models, which can also connect the different regions over broad ranges of $\mu_B$ and $T$. The challenge is that the QCD phase diagram is multidimensional. This means that, e.g., with regard to isospin, while matter studied in laboratory experiments is nearly isospin symmetric, neutron stars and their mergers are not (with the matter in supernova explosions being an exception, closer to laboratory conditions,~\cite{Prakash:1996xs,Pons:1998mm}). As a result, laboratory and astrophysics data occupy different layers along the additional isospin dimension of the phase diagram. Another relevant dimension is strangeness, which is produced in dense astrophysical matter but not in laboratory experiments, where the short timescales of energetic particle collisions do not allow weak processes to take place (although strange particle-antiparticle pairs can be created without producing \emph{net} strangeness). Another relevant dimension is the magnetic field~(\cite{Andersen:2021lnk}), which can affect relativistic heavy-ion collisions, some neutron stars (those with extremely strong magnetic fields are known as  \emph{magnetars},~\cite{Mereghetti:2015asa}), and neutron-star mergers,~(\cite{Most:2025kqf}). It should be kept in mind that magnetic fields of the order of $B\sim10^{18}~\rm{G}$ are required to significantly affect the dense- and/or hot-matter EoS.

Consequently, connecting the entire QCD phase diagram requires relativistic dense-matter models that cover a broad range of $\mu_B$ and $T$, in addition to isospin, strangeness, and possibly magnetic fields. Although a few models can cover up to $5$ dimensions, e.g., the CMF model and the Polyakov-Nambu-Jona-Lasinio (PNJL) model~(\cite{Peterson:2023bmr}), they remain the exception to that rule. An alternative approach  is to combine different theories and models within their respective domains of applicability. Examples include EoS repositories, such as the CompStar Online Supernovae Equations of State (CompOSE,~\cite{Typel:2013rza,CompOSECoreTeam:2022ddl,Dexheimer:2022qhn}), as well as infrastructures that already connect EoSs, such as the more-general Modular Unified Solver of the Equation of State (MUSES,~\cite{ReinkePelicer:2025vuh}), the heavy-ion-collisions-oriented Beam Energy Scan Theory (BEST~\cite{An:2021wof}, and the neutron-star-oriented Crust Unified Tool for Equation-of-state Reconstruction (CUTER,~\cite{Davis:2025nwz}).

On the low-energy experimental side, the main open questions concerning dense matter include a more accurate determination of nuclear saturation properties, such as the derivatives of the symmetry energy and the hyperon potentials. The derivatives of the symmetry energy are essential for understanding how the equation of state depends on isospin and how this dependence evolves with density; see~\cite{Sorensen:2023zkk} for details. The hyperon potentials determine densities at which hyperons appear in dense matter. Together, these quantities strongly affect the EoS predicted by any dense-matter model, although they cannot be uniquely determined from neutron-star observations because of the many degeneracies, particularly those associated with deconfinement to quark matter.
On the observational side, the main open questions include determining the maximum mass of neutron stars, together with obtaining more accurate radius measurements. In particular, there should exist a maximum-mass limit for (non-rotating, non-magnetized) neutron stars, analogous to the Chandrasekhar limit for white dwarfs. Recently, several \emph{gap} objects have been identified through both electromagnetic and gravitational-wave observations in the mass range $2-5~\rm{M}_{\rm{Sun}}$, laying between the standard ranges of masses of neutron star and black holes. These are the ones that were observed gravitationally: 
GW190814 with a mass $2.59^{+0.08}_{–0.09}~\rm{M}_{\rm{Sun}}$~(\cite{LIGOScientific:2020zkf}),
GW190917 with a mass $2.1^{+1.1}_{–
0.4}~\rm{M}_{\rm{Sun}}$~
(\cite{LIGOScientific:2021usb}),
GW200210 with a mass $2.83^{+0.47}_{–0.42}~\rm{M}_{\rm{Sun}}$~
(\cite{KAGRA:2021vkt}), and
GW230529 with a mass $3.6^{+0.8}_{–1.2}~\rm{M}_{\rm{Sun}}$~
(\cite{LIGOScientific:2024elc}).
Once these together with the gap objects observed gravitationally are conclusively classified, they will further constrain the maximum mass limit for neutron stars.

Most importantly, the main question that remains open on the high-energy side concerns the quark deconfinement phase transition: where does it take place in the dense region of the QCD phase diagram? Is there a critical point accompanied by a first-order phase-transition coexistence line? Currently, many laboratories around the world (such as the Facility for Antiproton and Ion Research, FAIR, in Germany) together with new telescopes and interferometers on Earth and in space, are being built to address these questions, with each new result bringing us one step closer to understanding how the Universe behaves at its most fundamental level.

\section*{Further reading}

\begin{itemize}
    \item {\bf{Books:}} \cite{Shapiro:1983du,Serot:1984ey,Glendenning:1997wn,Haensel:2007yy,Schmitt:2010pn,Schaffner-Bielich:2020psc,Ratti:2021ubw,Bandyopadhyay:2021gnp,ColemanMiller:2021lky,Benhar2023}
    \item {\bf{Collected volumes:}} \cite{Uechi:2012wzc,Rezzolla:2018jee,Blaschke2001PhysicsNSI}
    \item {\bf{Review papers:}} \cite{Lovato:2022vgq,Sorensen:2023zkk,MUSES:2023hyz}
\end{itemize}

\begin{ack}[Acknowledgments]

I would like to thank Rajesh Kumar and Rafael Jacobsen for providing figures for the chapter and Michael Mandl, Gert Aarts, the Astrophysics research group at Kent State University and the MUSES and NP3M collaborations for useful discussions. 
\end{ack}

\bibliographystyle{Harvard}
\bibliography{reference}

\end{document}